\documentstyle[prl,aps,multicol,epsfig]{revtex}

\begin{document}
\large

\title{Quantum Information Processing by NMR using strongly coupled spins}
\author{T.S. Mahesh$^{\dagger 1}$, Neeraj Sinha$^{\dagger 2}$, Arindam Ghosh$^{\dagger}$, Ranabir Das$^{\dagger}$, 
                    N. Suryaprakash$^{\ddagger}$ \\ 
         Malcom H. Levitt$^{\top}$, K.V. Ramanathan$^{\ddagger}$ and Anil Kumar$^{\dagger \ddagger *}$\\
        $^{\dagger}$ {\small \it Department of Physics, Indian Institute of Science, Bangalore, India}\\
        $^{\ddagger}$ {\small \it Sophisticated Instruments Facility, Indian Institute of Science, Bangalore, India}\\
        $^{\top}$ {\small \it Department of Chemistry, University of Southampton, Southampton SO17 1BJ, England}\\}

\maketitle
\hspace{7.5cm} {\small \bf ABSTRACT}

\begin{abstract}
The enormous theoretical potential of Quantum Information Processing (QIP) is driving the
pursuit for its practical realization by various physical techniques. Currently Nuclear Magnetic
Resonance (NMR) has been the forerunner by demonstrating a majority of quantum algorithms. In
NMR, spin systems consisting of coupled nuclear spins are utilized as qubits. In order to carry out
QIP, a spin system has to meet two major requirements: (i) qubit addressability and (ii) mutual
coupling among the qubits. It has been demonstrated that the magnitude of the mutual coupling
among qubits can be increased by orienting the spin-systems in a liquid crystal matrix and utilizing
the residual dipolar couplings. While utilizing residual dipolar couplings may be useful to increase
the number of qubits, nuclei of same species (homonuclei) might become strongly coupled. In strongly coupled
spin-systems, spins loose their individual identity of being qubits. We propose that even such
strongly coupled spin-systems can be used for QIP and the qubit-manipulation can be achieved by
transition-selective pulses. We demonstrate experimental  preparation of pseudopure states,
creation of maximally entangled states, implementation logic gates and  implementation of Deutsch-Jozsa (DJ)  algorithm 
in strongly coupled 2,3 and 4 spin systems. The energy levels 
of the strongly coupled 3 and 4 spin systems were obtained by using a Z-COSY experiment. 
\end{abstract}
\footnote{ Present address: {\small \it Francis Bitter Magnet Lab, Massachusetts Institute of Technology, Cambridge,
          Massachusetts 02139, USA}\\
    $^{2}$ Present address: {\small \it Department of Chemistry, Iowa state University, Iowa 50011-3111, USA}\\
  $^*$For correspondence (e-mail:{\it anilnmr@physics.iisc.ernet.in})}

\section{Introduction}
The theoretical success of exploiting the quantum nature of physical systems in certain information
processing tasks like prime factorization \cite{shor} and unsorted database search \cite{grover} has motivated the
pursuit for the practical realization of Quantum Information Processing (QIP) \cite{preskill,chuangbook,bouwmeester}. With the
demonstration of many quantum algorithms, Nuclear Magnetic Resonance (NMR) is now considered as
a suitable test-bed for QIP. One of the main challenges for the progress of NMR QIP 
is ``how to increase the number of qubits?". In this direction several attempts  are being made,
 such as (i) find molecules with different chemical shifts and J-couplings, and (ii) use of dipolar 
and quadrupolar couplings. This paper concentrates on one aspect: 
 ``how to use dipolar couplings among homonuclear spins?".  This problem is outlined in the following paragraphs.

In NMR, systems consisting of coupled spin-1/2 nuclei form qubits.
In order to carry out QIP, the spin-system has to meet two main requirements: qubit addressability
and mutual coupling among the qubits. In liquid state NMR using isotropic fluids, the qubit addressability is normally
provided by the differences in Larmor frequencies of the various spin-1/2 nuclei, while the mutual coupling is
normally provided by the scalar (J) coupling among the nuclei connected by covalent bonds.
 The Hamiltonian for a J-coupled spin system is \cite{ernstbook},
\begin{eqnarray}
{\mathcal H}&=&{\mathcal H}_Z+{\mathcal H}_J \nonumber \\
            &=& \sum_i \omega_i I_{iz} +\sum_{i,j (i<j)} 2\pi J_{ij}(I_{iz}I_{jz}+I_{ix}I_{jx}+I_{iy}I_{iy}) 
\end{eqnarray}
 where ${\mathcal H}_Z$ is the Zeeman Hamiltonian, and ${\mathcal H}_J$ is the coupling Hamiltonian. 
When $2\pi J_{ij} \ll \vert \omega_i-\omega_j \vert$, the system is said to be weakly coupled, 
and the Hamiltonian can be approximated to  \cite{ernstbook},
\begin{eqnarray}
{\mathcal H}= \sum_i \omega_i I_{iz} +\sum_{i,j (i<j)} 2\pi J_{ij}I_{iz}I_{jz}
\end{eqnarray}

For qubit addressability, all $\omega_i$  should be sufficiently dispersed and all $J_{ij}$ should be non-negligible
($>$ 1Hz) and unequal in magnitude. In such a circumstance each spin can be treated as a qubit and the 
coupled nuclei as several qubits. The values of $J_{ij}$ depend on the covalent bonds connecting spins i
and j, and normally has a small range ($< 10^2$ Hz) and becomes too small ($< 1$ Hz) if the spins are
connected by more than 4-5 covalent bonds. This places a natural limit on the number of qubits
reachable by liquid state NMR using J-couplings alone. To overcome this limitation the possibility
of using dipolar couplings was considered. The truncated Hamiltonian for dipolar interaction
is \cite{ernstbook,slic}, 
\begin{eqnarray}
{\mathcal H}_D= \sum_{i,j (i<j)} 2\pi D_{ij}(3I_{iz}I_{jz}-{\mathrm{\bf I_i.I_j}})
\end{eqnarray}
The dipolar coupling $D_{ij}$, between spins of gyromagnetic ratios $\gamma_i$ and $\gamma_j$ whose inter-distance vector
$r_{ij}$ makes an angle $\theta_{ij}$ with the Zeeman magnetic field is of the form \cite{ernstbook,slic},
\begin{eqnarray}
D_{ij}= \frac{\gamma_i \gamma_j \hbar}{4 \pi r^3_{ij}}(1-3cos^2\theta_{ij})
\end{eqnarray}

Dipolar couplings among common nuclear species have larger magnitudes ($\sim$ $10^3$ Hz) and longer
range than the scalar couplings. However, in isotropic liquids the time average of $D_{ij}$ vanishes, while
in solids there are too many dipolar couplings resulting in broad unresolved lines and loss of qubit
addressability. In molecules oriented in a liquid crystal matrix, while the intermolecular dipolar
couplings are vanishingly small, the intra molecular dipolar couplings survive, scaled down by the
order parameter $(S_{ij})$ of the liquid crystal \cite{slic,khetrapal},
\begin{eqnarray}
D_{ij}^{ori}= \frac{\gamma_i \gamma_j \hbar}{4 \pi r^3_{ij}} < 1-3cos^2\theta_{ij} >
=- \frac{\gamma_i \gamma_j \hbar}{4 \pi r^3_{ij}} S_{ij}
\end{eqnarray}
In such systems one obtains a finite number of sharp well resolved spectral lines making it possible
to use such systems for NMR-QIP. In the NMR-QIP experiments implemented so far, the
systems have been chosen such that either (i) $2\pi(J_{ij} + 2D_{ij}) \ll \vert (\omega_i - \omega_j) \vert$, yielding weakly coupled
spin-systems which is the case for the heteronuclear spin-systems \cite{chuang00b,marka} or (ii) the coupling
$2(J_{ij} + 2D_{ij})$ is finite and $\vert \omega_i - \omega_j \vert$ = 0, i.e., 
equivalent-spins case \cite{fung00,mahesh02}. In the latter case, the
symmetry filtering of energy levels become increasingly difficult for higher number of qubits \cite{mahesh02}. 
Even though the heteronuclear spins oriented in liquid crystal matrix are excellent for QIP since
they provide very good qubit addressability as well as large mutual
coupling, the use of more than 3 to 4 heteronuclear spins is
limited by the extensive hardware requirements.  Therefore for reaching larger
number of qubits, one needs to utilize homonuclear (nuclei of same species having same $\gamma$
 but different chemical shifts ) spins oriented in a liquid crystal.

Homonuclear spins oriented in a liquid crystal generally become strongly coupled since 
the dipolar couplings become comparable to or more than the differences in Larmor frequencies $\vert \omega_i - \omega_j \vert$.
In such a situation, the Zeeman and the coupling parts of the Hamiltonian do not commute. Therefore the eigenstates of strongly 
coupled spins are obtained as the linear combinations of product states of various spins 
and the individual spins can no more be treated as qubits.
We propose and demonstrate here that, the $2^{N}$ eigenstates of a coupled N-spin 1/2  system
can be treated as an N qubit system even in the presence of strong coupling. Similar idea has already been used in demonstration 
 of QIP using quadrupolar (S $>$1/2) nuclei oriented in high magnetic field, where the 2S+1 non-equidistant energy levels have been 
 treated as  N-qubit systems, where $2^N=2S+1$. So far S=3/2 and 7/2  have been utilized respectively as 2 and 3 qubit systems 
\cite{fun,mulf,neeraj01,mur,ranapra}, for various NMR-QIP. 

While substantial work has been carried out in NMR-QIP using weekly coupled spin systems \cite{jonescrit}, 
till now the use of the strongly coupled spin systems for QIP has not been experimentally demonstrated, presumably because 
(i) spin-selective pulses are not defined in the case of strongly coupled spins
\cite{ernstbook}, and (ii) the difficulty in constructing a general unitary operator using the evolution under scalar 
coupling
\cite{cory98b}.  
The problem of using scalar coupling evolution of a strongly coupled two-spin system for a general unitary transform has recently 
been addressed theoretically, but extending to a N-spin system is complicated \cite{benjamin}.  However, we note that unlike spin-selective
pulses, the transition-selective pulses are well defined even in strongly coupled spin-systems \cite{ernstbook} and hence it is possible 
to construct a unitary transformation using transition selective pulses. In section \ref{2spinstrong} we demonstrate NMR-QIP on
a strongly coupled two spin-system in isotropic medium by preparing pseudopure states, implementing DJ algorithm, by creating 
Einstein-Podolsky-Rosen (EPR) state and by implementing logic gates.  Section \ref{3spinstrong}
describes the creation of Greenberger-Horne-Zeilinger (GHZ) states and implementation of two-qubit DJ algorithm on a 
strongly coupled three spin 
system in an oriented medium
after labeling the transitions using the Z-COSY experiment. Labeling of transitions, preparation of pseudopure states,
and implementation of gates on a four-spin strongly coupled system are demonstrated in section \ref{4qstrong}.  

 In this study the strongly coupled systems used are:
\begin{center}

\begin{tabular}{|c|c|c|c|c|}
\hline
Sl. No.& Sample & Solvent & Isotropic/ Oriented & No. of qubits \\ 
\hline 
\hline 
1 & Trisodium citrate & $D_2$O & Isotropic & 2 \\
\hline
2 & Organometallic compound (I)  & CDCl$_3$ & Isotropic & 2 \\
\hline
3 & 1-bromo-2,3-dichlorobenzene & ZLI-1132 & Oriented & 3 \\
\hline
4 & 2-chloroiodobenzene & ZLI-1132 & Oriented & 4 \\
\hline
\end{tabular} 

\end{center}
The experiments have been carried out on a
Bruker DRX-500 NMR spectrometer at 300K temperature.

\section{Two-spin strongly coupled system}
\label{2spinstrong}
The four eigenstates of a strongly coupled 
two-spin system (of spin 1/2 nuclei; AB spin-system)
in isotropic medium are,
$\vert \alpha \alpha \rangle$,
$\cos\Theta \vert \alpha \beta \rangle+\sin\Theta \vert \beta \alpha \rangle$,
$\cos\Theta \vert \beta \alpha \rangle-\sin\Theta \vert \alpha \beta \rangle$, and
$\vert \beta\beta \rangle$ (Figure 1(a)), where
$\Theta=\frac{1}{2}\tan^{-1}(2 \pi J_{AB}/(\omega_A-\omega_B))$
\cite{ernstbook}.
These eigenstates are labeled respectively as
$\vert 00 \rangle$,
$\vert 01 \rangle$,
$\vert 10 \rangle$, and
$\vert 11 \rangle$,  thus forming a two-qubit system 
(Figure 1(b)).
To demonstrate QIP on such a system, we have taken the strongly
coupled $^1$H spins of trisodium citrate (Figure 1(c)).
In this system, the scalar coupling (J) is 15 Hz,
the difference in Larmor frequencies ($\Delta f$) is 55.5 Hz, and
the strong coupling parameter ($\Theta$) is $7.6^\circ$.
The equilibrium spectrum of the system is shown 
Figure 1(d).

\subsection{Preparation of Pseudopure states}
In QIP, the computation normally begins from a 
definite initial state known as a pure state
\cite{preskill,chuangbook}.
In NMR however because of the small energy gaps, it is not possible 
to realize a pure state, wherein the whole population is in one energy level, since it requires very low temperatures as well as
very high magnetic fields.  However, an alternate solution was discovered to overcome this problem \cite{cory97,chuang97}. 
In thermal equilibrium, NMR density matrix can be written as 
\begin{equation}
\rho=2^{-N} \{I + \epsilon \rho_{dev}\}.
\end{equation}
The first part is a normalized unit matrix which corresponds to a uniform population background.  
The second part containing the traceless deviation density matrix $\rho_{dev}$ (with a small coefficient 
$\epsilon \sim 10^{-5}$) evolves under various NMR Hamiltonians, and gives measurable signal.  
It was observed independently by Cory et al \cite{cory97} and Chuang et al \cite{chuang97}, 
that by applying certain pulse sequence to the system in equilibrium,
we can prepare the so called pseudopure density matrix,
\begin{equation}
\rho_{pps}=2^{-N} \{(1-\epsilon'/2^N)I + \epsilon' \rho_{pure}\}.
\end{equation}
The first part is again a scaled unit matrix, but the second
part corresponds to a pure state. Such pseudopure states mimic pure states \cite{cory97,chuang97}.
Many methods have been proposed for the preparation of pseudopure states including 
spatial averaging \cite{cory97,cory98}, temporal averaging \cite{knill98},
logical labeling \cite{chuang97,chuang98,kavita00}, and spatially averaged logical labeling \cite{maheshpra01}.
Some of the other methods include the preparation of pseudopure states via cat states \cite{knillnature} and
preparation of pair of pseudopure states \cite{fung00}. 
Both J-evolution and transition selective pulse methods have been utilized for preparation 
of pseudopure states \cite{cory97,knill98,ernst,kavita00,neeraj01,peng}.

 We have adopted the method of spatial averaging using transition selective pulses.
The Boltzmann distribution of populations at high-temperature 
approximation is linear with energy gap. The equilibrium populations (in excess of a large uniform background
 population) of a homonuclear 
two spin system is given in Fig 2(a). For creating a pseudopure state all the populations except 
one of the states have to be equalized.  This distribution can be achieved by 
a sequence of transition selective pulses intermittent with field gradient pulses to destroy any coherence created in the process.

A transition-selective pulse of nutation angle
$\theta$ and of any transverse phase between states $(i,j)$,
changes the populations as follows:
\begin{eqnarray}
p_i^\prime &=& p_i \cos^2(\theta/2)+p_j \sin^2(\theta/2) \nonumber \\
p_j^\prime &=& p_j \cos^2(\theta/2)+p_i \sin^2(\theta/2).
\label{popurf}
\end{eqnarray}
 To prepare the $\vert 00 \rangle$ pseudopure state (Fig 2(b)) from equilibrium (Fig 2(a)), we use a sequence 
$[(\theta)^{\vert 10 \rangle \leftrightarrow \vert 11 \rangle}-G_z
-(90^\circ)^{\vert 01 \rangle \leftrightarrow \vert 11 \rangle}-G_z^{\prime}]$ (pulses are applied from left to right), 
 and it is inferred from the deviation populations of Fig 2(a) and (b), that $\theta$ should be such that 
$\cos^2(\theta/2)$=2/3. This yields $\theta=70.5^\circ$. $\theta=90^\circ$ on the other hand equalizes the 
populations of the two levels to an average value.
 Thus, the $\vert 00 \rangle$ pseudopure state is prepared by
the pulse sequence $[(70.5^\circ)^{\vert 10 \rangle \leftrightarrow \vert 11 \rangle}-G_z
-(90^\circ)^{\vert 01 \rangle \leftrightarrow \vert 11 \rangle}-G_z^{\prime}]$ and the corresponding spectrum is given in Fig. 2(b).
The $\vert 01 \rangle$ and $\vert 10 \rangle$ pseudopure  states (Figure 2(c), 2(d)) 
are respectively prepared by applying respectively $(180^\circ)^{(\vert 00 \rangle \leftrightarrow \vert 01 \rangle)}$ 
and  $(180^\circ)^{(\vert 00 \rangle \leftrightarrow \vert 10 \rangle)}$ pulse after creating the 
$\vert 00 \rangle$ pseudopure state.  The $\vert 11 \rangle$ pseudopure state (Figure 2(e)) is prepared by the pulse sequence 
$(70.5^\circ)^{(\vert 00 \rangle \leftrightarrow \vert 01 \rangle}-
G_z-(90^\circ)^{\vert 00 \rangle \leftrightarrow \vert 10 \rangle}-G_z^{\prime}$. The observed intensities in 
the spectra on the left hand side of Fig. 2 correspond to the created population distribution and hence confirm the creation 
of the pseudopure states.

\subsection{Deutsch-Jozsa Algorithm}
Deutsch-Jozsa (DJ) algorithm is one of the first quantum algorithms which 
successfully demonstrated the power of QIP \cite{deu,cleve}.  The task of DJ algorithm is
to distinguish between two classes of 
many-input-one-output functions, constant and balanced. Constant functions are those functions in which  all the outputs are
same independent of inputs; and  balanced functions are those in which half the number of
inputs give one output and the other half gives another output.  Classically,
given a function of n input bits, it takes $2^{n-1}+1$ function-calls
on an average, to
determine whether the function is constant or balanced, whereas DJ 
algorithm needs only one function-call for any number of qubits.  DJ algorithm
has been implemented in NMR using scalar coupling evolution as well as  using
spin and transition selective pulses \cite{kavita00,djchu,djjo,free}.  We have followed Cleve's version 
of DJ algorithm which requires one extra work qubit \cite{cleve}.
The circuit diagram and the NMR pulse sequence for implementing 1-qubit DJ algorithm
are shown in Figures 3(a) and 3(b) respectively.  The experiment begins
with the $\vert 00 \rangle$ pseudopure state. An initial $(\pi/2)_{-y}$ pulse (the pseudopure Hadamard operation 
 \cite{djjo}) on all the qubits creates
a superposition.  It may be noted that unlike weakly coupled spins the superposition created here 
is not uniform in the eigenbasis, since the coefficients of various eigenstates are different.
 However, as is shown here it is still possible to distinguish between the different classes of functions. 
 The Hadamard operation is followed by an unitary operator $U_f$ corresponding to
the given function $f$.  The unitary operator carries out the transformation
$\vert r,s \rangle \stackrel{U_f}{\rightarrow} \vert f(s) \oplus r,s \rangle$,
where $\vert r \rangle $ and $\vert s \rangle $ are the states of the work-qubit and the input-qubit
respectively.  
The four different one-qubit functions and corresponding unitary operator as well as r.f. pulses are listed in Table 1.  The
test for a balanced function is that the transitions of input-qubit will gain opposite phases at the end of the algorithm.
The experimental results corresponding to all the four functions $f_1$, $f_2$, $f_3$, and $f_4$ 
along with their corresponding simulated spectra are given in Figures 3(c)-3(f).  
From the spectra we can identify that functions  $f_1$ and  $f_2$ are constant since the transitions of 1 and 2 are of same phase, 
whereas  $f_3$ and $f_4$ are balanced since the transitions of 1 and 2 are of opposite phase.
\subsection{Creation of an EPR state}

Einstein-Podolsky-Rosen (EPR) pairs are the maximally entangled pairs
of the form 
\begin{eqnarray}
&& \frac{1}{\sqrt{2}}(\vert 00 \rangle \pm \vert 11 \rangle) \;\;\;\; 
{\mathrm {or}} \nonumber \\
&& \frac{1}{\sqrt{2}}(\vert 01 \rangle \pm \vert 10 \rangle)
\end{eqnarray}
which are not reducible into product states of individual qubits \cite{preskill}.  The non-local 
correlation exhibited by these pairs have no classical equivalence, and are exploited in many branches
of quantum information processing including quantum computation and quantum teleportation \cite{preskill,chuangbook}.
EPR states of a pair of weakly coupled nuclear spins have been earlier created by NMR using
spin selective pulses and evolution of coupling \cite{chuang00b}.  

Here we demonstrate the creation of EPR state on the above strongly coupled two-spin system using transition selective pulses
and tomograph the result using  non-selective pulses. Starting from $\vert 00 \rangle$ pseudopure state, 
the EPR state $(\vert 00 \rangle + \vert 11 \rangle)/\sqrt{2}$ can be created by applying the pulse sequence 
(pulses are to be applied from left to right)
\begin{equation}
\left [ 
\left (\frac{\pi}{2} \right )^k_{\phi_1}
\cdot
(\pi)^l_{\phi_2}
\right ],
\label{eprpulex}
\end{equation}
where $k$ and $l$ are the transitions, and $\phi_1$ and $\phi_2$
are the phases as shown in Table 2.  For example,
the unitary operator U for the pulse sequence 
$\left [
\left (\frac{\pi}{2} \right )^2_x
\cdot
(\pi)^3_{-x}
\right ]
$ is
\begin{eqnarray}
U &=& \exp \left ( i  \pi 
I_{x}^{\vert 10 \rangle \leftrightarrow \vert 11 \rangle}
\right )
\cdot
\exp \left ( -i \frac{\pi}{2}
I_{x}^{\vert 00 \rangle \leftrightarrow \vert 10 \rangle}
\right ) \nonumber \\
&=& \frac{1}{\sqrt{2}} \left (
\begin{array}{cccc}
1 & 0 & -i & 0 \\
0 & \sqrt{2} & 0 & 0 \\
0 & 0 & 0 & i\sqrt{2} \\
1 & 0 & i & 0 \\
\end{array}
\right ),
\end{eqnarray}
 where the operators are however applied from right to left.
EPR state is obtained by applying $U$ on $\vert 00 \rangle$
pseudopure state
\begin{eqnarray}
U \cdot 
\pmatrix{1 & 0 & 0 & 0 \cr 0 & 0 & 0 & 0 \cr 0 & 0 & 0 & 0 \cr 0 & 0 & 0 & 0} 
\cdot U^{\dagger} =
\frac{1}{2}
\pmatrix{1 & 0 & 0 & 1 \cr 0 & 0 & 0 & 0 \cr 0 & 0 & 0 & 0 \cr 1 & 0 & 0 & 1}
\end{eqnarray}
  A phase-cycle over different combinations given in Table 2  
helps to reduce errors in the off-diagonal elements.
Figure 4(a) shows the experimental equilibrium spectrum and 4(b) shows 
the experimental spectrum after creating 
the EPR state.  Since EPR state does not consist of any single quantum coherence, no signal is obtained
(Figures 4(b)). The corresponding simulated spectrum are also shown on the right hand side in each figure.   

To verify the creation of the EPR state it is necessary to tomograph the complete density matrix.
Tomography in NMR is normally carried out using spin-selective pulses
obtaining a series of one-dimensional NMR experiments each giving a linear equation of different
elements of the density matrix \cite{chuang98}.  However, in the case of strongly coupled systems, all operations
including tomography excludes the use of spin selective pulses and demands either non-selective pulses or
transition selective pulses or both. Recently, a robust method for tomography was suggested based on
two-dimensional Fourier spectroscopy, which utilizes only non-selective pulses \cite{ranabirtomo}. 
This method involves:\\ 
 (i) an one-dimensional experiment for measuring 
diagonal elements:  $ [ G_z - 10^\circ_x-t]$,  and  \\
(ii) a two-dimensional multiple quantum experiment
for measuring all off-diagonal elements: $[t_1-(\pi/2)_y-G_z-(\pi/4)_{-y}-t_2]$.
\begin{eqnarray}
\end{eqnarray}
The scheme involves only non-selective pulses and therefore it is not only
simple and accurate but also applicable to strongly coupled
systems. The result of measurement of diagonal elements of the EPR state [experiment (i)] is shown in Figure 4(c) 
and the corresponding simulated spectrum is shown in Figure 4(h).
  Figure 5
shows the complete pulse-sequence for creation of EPR state followed by measurement of the off-diagonal elements [experiment (ii)]. 
The resulting 2D spectrum of experiment is given in 4(k), which clearly shows the double quantum peaks 
corresponding to the EPR state. No zero quantum or single quantum peaks are observed.  
\cite{ranabirtomo}.

Since the diagonal and off-diagonal terms are measured by two different schemes, it is necessary also to determine
the scaling between the two measurements. Normally this is achieved by comparing single quantum terms
in the two-dimensional experiment with the spectrum obtained by direct detection of the density
matrix  \cite{ranabirtomo}.  However since no single quantum coherence is present in the EPR density matrix, we
carried out two additional one-dimensional experiments: (iii)[ $(\pi/4)_y-t$] and (iv) [$(\pi/4)_x-t$] after creation of EPR pair. 
 Experimental spectra corresponding to (iii) and (iv) are shown in Figures 4(d) and
4(e) respectively, and corresponding simulated spectra are shown in Figures 4(i) and 4(j)
respectively.  The signals in 4(d) are proportional to the sum of the amplitudes of diagonal and double quantum coherences 
of the EPR state while those in 4(e) are proportional to the differences (when single quantum coherences are not present 
as in the present case). Since in a perfect EPR state the diagonal elements and double quantum coherences are equal, 
the spectrum of Fig 4(e) should have no signal as is evident from the simulated spectrum of Fig. 4(j). The signals of 
Fig. 4(e) compared to Fig 4(d) are measures of experimental errors, which in the present case are estimated to be less than 
15$\%$ \cite{mahthesis}. The complete density matrices corresponding
to the theoretical and experimentally obtained EPR state are shown in Figures 4(l) and 4(m).

\subsection{Implementation of Logic gates}
  Logic gates have been implemented earlier by one and two-dimensional NMR 
 using weakly coupled spin-1/2 nuclei as well as quadrupolar nuclei \cite{cory98,chuang97b,ernst,mahesh01,neeraj01}. 
We demonstrate here the first implementation of a complete set of 24 one-to-one 
logic gates in a 2-qubit system using strongly coupled spin-1/2 nuclei. The system chosen for 
this purpose is the two phosphorus nuclei of the organometallic compound (I) shown in Fig. 6(a).
 The energy level diagram (Fig. 6(b)) and the  equilibrium phosphorus spectrum of this molecule in 
isotropic liquid state is given in Fig. 6(c). Starting from 
equilibrium, the logic gates were implemented  using sequences of transition selective pulses.    
 The final populations were mapped by a small angle $(10^o)$ non-selective pulse.  
The spectra corresponding to final populations of all the 24 logic gates are given in Fig 7. The unitary transforms and 
pulse sequences for implementation of these gates are given in reference \cite{mahesh01} with the modification that the r.f. power 
has been adjusted for given angle of flip for the two inner versus the two outer transitions.

\section{Three-spin strongly coupled system}
\label{3spinstrong}
\subsection{The system and labeling of transitions}
The system chosen is the three strongly coupled protons of 3-bromo-1,2-dichlorobenzene (Figure
8) oriented in the nematic liquid crystal ZLI-1132.  The equilibrium spectrum of the system
at 300 K obtained from DRX 500 MHz spectrometer is shown in Figure 8.  There are only nine 
out of a total 15 possible single quantum transitions with observable intensity, in this spin system.  Construction of energy
level diagram and labeling of transitions were performed using a Z-COSY experiment \cite{zcosy1,zcosy2}. 
 The Z-COSY spectrum along with  cross-sections  parallel to $\omega_2$ axis
at various transitions  is given in Figure 9.
The zero-quantum artifacts were suppressed in the Z-COSY
experiment by incrementing a delay synchronized with t$_1$ increment \cite{zcosy1,zcosy2}.
The connectivity matrix is obtained by the MATLAB assisted automation
\begin{eqnarray}
\begin{array}{r|c}
&
\begin{array}{rrrrrrrr}
\;\;\;\;1\;\;\; & 2\;\;\; & 3\;\;\; & 4\;\;\; & 5\;\;\; & 6\;\;\; & 7\;\;\; & 8\;\;\;
\end{array}
\\
\hline
\\
\begin{array}{c}
1 \\
2 \\
3 \\
4 \\
5 \\
6 \\
7 \\
8
\end{array}
&
\left [
\begin{array}{rrrrrrrr}
0 & 0 & -1 & 1 & 0 & 0 & 1 & 0 \\
0 &0 & 1 & -1 & -1 & 0 & 0 & 1 \\
-1 & 1 &0 & 0 & 1 & 0 & 0 & 0 \\
1 & -1 & 0 &0 & 0 & 0 & -1 & 1 \\
0 & -1 & 1 & 0 &0 & 1 & -1 & 0 \\
0 & 0 & 0 & 0 & 1 &0 & 1 & -1\\
1 & 0 & 0 & -1 & -1 & 1 &0 & 0 \\
0 & 1 & 0 & 1 & 0 & -1 & 0 & 0
\end{array}
\right ]
\end{array}.
\end{eqnarray}
The constructed energy level diagram for the above connectivity
matrix is shown in Figure 10. The ninth transition shown by dashed line belongs to the transition 011$\leftrightarrow$100, and is not 
connected to any other observed transitions. Therefore the transition did not show any connectivity to other transitions in Z-COSY 
experiment (Figure 9) and is marked as * in Figure 8. It turns out that these nine transitions are sufficient to carry out 
certain QIP operations as shown in section (B),(C) and (D).

\subsection{Preparation of Pseudopure state}
 We have used the method of ``POPS" to prepare a pair of  pseudopure states on this three-spin strongly coupled system \cite{fung00}. 
   POPS requires only two population distributions:  (i) Equilibrium  populations (Fig. 11(a)) and  
(ii) Equilibrium populations changed by a single transition selective $\pi$ pulse on a given transition (Fig. 11(b)). 
Subtraction of (ii) from (i) yields effectively a pair of pseudopure states 
$\vert 000 \rangle \langle 000 \vert - \vert 001 \rangle \langle 001 \vert$ (Fig. 11(c)). 

\subsection{Creation of $\vert {\sf GHZ} \rangle \langle
{\sf GHZ} \vert - \vert 001 \rangle \langle 001 \vert$ state}
\label{ghzcreation}
Entanglement between many particles is essential for most quantum
communication schemes, including error-correction schemes and
secret key distribution network \cite{bouwmeester}.
Greenberger-Horne-Zeilinger (GHZ) states are three-spin
entangled states of the form \cite{ghz1,ghz2}
\begin{equation}
\vert {\sf GHZ} \rangle = \frac{1}{\sqrt{2}}
\left(
\vert 000 \rangle + \vert 111 \rangle
\right )
\end{equation}
Three particles in GHZ state exhibit one of the strangest
correlations that can not be explained by any hidden variable
theory \cite{ghz1,ghz2}.  A set of measurements carried out on
three particles in GHZ state prepare the particles in a
classically impossible correlated state \cite{bouwmeester}.
In NMR, the GHZ state was first created by Laflamme et al
\cite{laflamme98}.  The correlations of the GHZ state has
been studied using NMR by Nelson et al \cite{nelson00}.

The preparation of GHZ state requires preparing a pseudopure initial state, like $\vert 000 \rangle$. 
 However, since we have prepared pairs of pseudopure states as the
initial state, we will be actually preparing a state
\begin{equation}
\vert {\sf GHZ} \rangle \langle {\sf GHZ} \vert -
\vert 001 \rangle \langle 001 \vert.
\end{equation}
This state differs from GHZ state only in the diagonal
elements and therefore retains the essential correlations of the
GHZ state.

The GHZ state can be created from the $\vert 000 \rangle$
pseudopure state using a cascade of three transition selective pulses
(i) $(\pi/2)_{\phi_1}$ pulse on the transition {\bf 8}, (ii) $(\pi)_{\phi_2}$ pulse
on the transition {\bf 4} and (iii) $(\pi)_{\phi_3}$ pulse on the transition 
{\bf 1}.  The pulse sequence is,
\begin{eqnarray}
[ \left ( \frac{\pi}{2} \right )_{\phi_1}^{(8)} \cdot 
\pi_{\phi_2}^{(4)} \cdot \pi_{\phi_3}^{(1)} ], 
\label{ghzpulses}
\end{eqnarray}
 where pulses are to be applied from left to right.
The phases $\phi_1$, $\phi_2$ and $\phi_3$ of the pulses 
can be any one of the 16 possible combinations, as shown in Table 3.  A phase-cycle over these combinations helps to
reduce the errors in the off-diagonal elements. \\
The unitary operator for the above pulse-sequence can be written as,
\begin{eqnarray}
U_{\sf GHZ} &=&
\exp(-i \pi I^{(1)}_{\phi_3})
\cdot
\exp(-i \pi I^{(4)}_{\phi_2})
\cdot
\exp\left (-i \frac{\pi}{2}I^{(8)}_{\phi_1} \right ),
\end{eqnarray}
where $I^{(1)}_{\phi_3}$, $I^{(4)}_{\phi_2}$ and $I^{(8)}_{\phi_1}$
are the single transition operators on transitions {\bf 1},
{\bf 4} and {\bf 8} respectively.  Spectrum corresponding to the
diagonal part of $\vert {\sf GHZ} \rangle \langle {\sf GHZ} \vert
- \vert 001 \rangle \langle 001 \vert$ state is shown in Figure
11(d). It is clearly seen that in the GHZ state (Eq. 15) transitions 1,3 and 8 have approximately 
half the intensity of the equilibrium spectrum. Transitions 5,6 and 7 appear due to POPS. However to confirm the creation 
 of GHZ state a complete tomography of the created GHZ state is needed. 
The pulse sequence for preparation of POPS, creation of 
$\vert {\sf GHZ} \rangle \langle {\sf GHZ} \vert- \vert 001 \rangle \langle 001 \vert$ state
 followed by tomography (using experiment (ii) of expression (13)) 
is given in the Figure 12(a).   
The 2D spectrum corresponding to the measurement of all off-diagonal elements of  
GHZ state is shown in Figure 12(b).  
Presence of only triple quantum coherence and absence of all other coherences confirms the creation of GHZ state. 
 The axial peaks at zero frequency are $\Omega_1$ dimension
is due to the longitudinal relaxation during the $t_1$ period and imperfections of the $90^o$ pulse (see caption of Fig. 4). 

\subsection{Two-qubit DJ using two-dimensional NMR}
Implementation of DJ-algorithm on two qubits requires three qubits including one work qubit. The algorithm can be described as

\begin{equation}
\vert r \rangle \vert s \rangle \vert t \rangle \stackrel{U_f}{\rightarrow} \vert r \rangle \vert s \rangle \vert t\oplus f(r,s) \rangle
\end{equation}
$\vert r \rangle \vert s \rangle$ and $\vert t \rangle$ being the states of the two input qubits ($I_1$,$I_2$)
and the work qubit ($I_0$) respectively. There are eight possible
two-bit binary functions (f) of which two are constant and six balanced.
The transformations corresponding to the constant functions $f_1$ and $f_2$ are respectively unity operator and a $\pi$ pulse on all the
transitions of the work
qubit. The unitary transformations encoding the remaining six balanced functions $f_3-f_8$ are acheived by transition selective
pulses on different transitions (3,4,9(*),6) of the work qubit as $[0,0,\pi,\pi]$,$[\pi,\pi,0,0]$,$[\pi,0,\pi,0]$,$[0,\pi,0,\pi]$,
$[\pi,0,0,\pi]$ and $[0,\pi,\pi,0]$;\cite{mahesh01}. \\
The pulse sequence for two-dimensional DJ algorithm is $[(\frac{\pi}{2})^{I_0 ,I_1 ,I_2} - t_1 - U_f - Det (t_2)]$.
The transitions of $I_1$ and $I_2$ qubits are frequency labelled
during the $t_1$ period and detected during the $t_2$ period. Fourier transformation with respect to $t_1$ and $t_2$ yields the desired 
two-dimensional spectrum \cite{mahesh01}.\\
The experimental result of the above operations on the strongly coupled 3-qubit system of 3-bromo-1,2-dichloro benzene is given in Figure
13. The experimental
results match the expected theoretical results, confirming that two-dimensional DJ algorithm can also be carried out in a strongly coupled
three spin system.

\section{Strongly coupled 4-spin system}
\label{4qstrong}
\subsection{Labeling of transitions}
The system chosen is 2-chloroiodobenzene dissolved in ZLI-1132.  The four aromatic protons form a strongly
coupled 4-spin system. Equilibrium spectrum is shown in Figure 14.  The transitions
are labeled according to the descending order of their intensities.  The total
number of transitions for a 4-spin strongly coupled system
is $^{2\times4}C_{4-1} = 56$ of which only 30 transitions have been observed with sufficient intensity. 
 There are other transitions of smaller intensity comparable
to that of $^{13}$C satellites. In order to avoid any interference due to $^{13}$C satellites we have used 
$^{13}$C decoupling in the experiment. Decoupling in $t_1$ dimension
is achieved by a single $\pi$ pulse on $^{13}$C channel in the middle
of $t_1$ period while that in $t_2$ dimension is carried out by multi-pulse decoupling sequences on the $^{13}$C channel.
The Z-COSY spectrum obtained after $^{13}$C decoupling for the present system is shown in Figure 15.  
The spectrum consists of more than 2000 peaks. The MATLAB analysis of the spectrum is carried out.
Figure 16(a) shows the connectivity and labeling of the 30 transitions of Figure 14.  

\subsection{4-Qubit gates and pseudopure states}
\label{4qstrongcnot}
Figure 16 shows implementation of gates and preparation of pairs of pseudopure
states using the 4-qubit system 2-chloroiodobenzene. The labeling scheme for the energy levels is shown
in Figure 16(a).  Figure 16(b) shows the equilibrium spectrum.  A C$^3$-{\sf NOT} gate can
be implemented using a single $\pi$ pulse on the transition 4,  The spectrum corresponding to the C$^3$-{\sf NOT} gate
obtained using a small angle detection pulse is shown in Figure 16(c).  A pair of pseudopure states, namely,
$\vert 1111 \rangle \langle 1111 \vert - \vert 1110 \rangle \langle 1110 \vert$
is prepared by subtracting the spectrum 16(c) from the equilibrium spectrum shown in 16(b).
The resultant spectrum is shown in Figure 16(d). Similarly the pair of pseudopure states
$\vert 1110 \rangle \langle 1110 \vert - \vert 1010 \rangle \langle 1010 \vert$
(Figure 16(e)) is prepared by inverting the transition 1 and subtracting
the obtained spectrum from the equilibrium spectrum. Figure 16(f) demonstrates the implementation
of C$^2$-{\sf SWAP} gate after preparing the pair of pseudopure states $\vert 1110 \rangle \langle 1110 \vert
- \vert 1010 \rangle \langle 1010 \vert$. The action of C$^2$-{\sf SWAP} gate is to interchange the
states $\vert 1110 \rangle$ and $\vert 1101 \rangle$.  This is achieved by three
transition selective pulses
\begin{eqnarray}
\left [
\pi^{(4)} \cdot \pi^{(14)} \cdot \pi^{(4)}
\right ],
\label{ccswapex}
\end{eqnarray}
where the superscripts indicate the transition numbers.
Since the states $\vert 1101 \rangle$ and $\vert 1110 \rangle$
have almost same populations in the present case, the spectrum
after applying C$^2$-{\sf SWAP} gate on equilibrium input state
will not be very much different from the equilibrium spectrum.
However, if one starts with a pair of pseudopure
states $\vert 1110 \rangle \langle 1110 \vert
- \vert 1010 \rangle \langle 1010 \vert$ as the input,
the output will be different pair of pseudopure states,
\begin{eqnarray}
\vert 1110 \rangle \langle 1110 \vert
- \vert 1010 \rangle \langle 1010 \vert
\stackrel{{\mathrm C}^2-{\sf SWAP}}{\longrightarrow}
\vert 1101 \rangle \langle 1101 \vert
- \vert 1010 \rangle \langle 1010 \vert.
\end{eqnarray}
The spectrum corresponding to the state
$\vert 1101 \rangle \langle 1101 \vert
- \vert 1010 \rangle \langle 1010 \vert$
obtained by applying C$^2$-{\sf SWAP} gate
on $\vert 1110 \rangle \langle 1110 \vert
- \vert 1010 \rangle \langle 1010 \vert$
is shown in Figure 16(f). The pair of pseudopure state
$\vert 1101 \rangle \langle 1101 \vert
- \vert 1010 \rangle \langle 1010 \vert$
can also be prepared by inverting the transition 15 and subtracting the spectrum obtained from
the equilibrium spectrum, as shown in Figure 16(g). The spectra in figures 16(f) and 16(g)
match fairly well, indicating good implementation of the C$^2$-{\sf SWAP} in Figure 16(f).

\section{Conclusions}
Increasing the number of qubits in NMR calls for the use of the dipolar couplings in 
oriented homonuclear systems which are generally strongly
coupled.  Spin-selective pulses are not defined in the
case of strongly coupled systems and qubit-addressability in such a scenario is achieved 
through transition selective pulses. It has been demonstrated earlier on weakly coupled systems,
that using only non-selective pulses and transition-selective pulses one can implement logic gates
\cite{kavita00,mahesh01,cory98b}, and algorithms such as Grover's algorithm and Quantum Fourier 
Transform \cite{ranajmr}. Efforts are ongoing to implement these algorithms in such systems as well as 
 to realize higher qubit systems using nuclear spins oriented in liquid crystal matrices.  
\section{Acknowledgments}
 The authors thank Dr. Swadhin K. Mandal and Prof. S.S. Krishnamurthy of the Department of 
Inorganic and Physical Chemistry, Indian Institute of Science, for the organometallic compound (I). 
The use of DRX-500 NMR spectrometer funded by the Department of Science and Technology, New Delhi, at the Sophisticated
Instuments Facility, Indian Institute of Science, Bangalore, is also gratefully acknowledged.


\newpage

\hspace{8cm} {\bf TABLES}\\\\
\begin{center}
Table 1. Functions, unitary operators and r.f. pulses for one-qubit DJ algorithm.\\
\begin{tabular}{|c|c|c|c|c|} \hline \hline
 &   \multicolumn{2}{c}{\ \ Constant} \quad \quad &  \multicolumn{2}{|c|}{\ \ \ Balanced}\\ \hline \hline
 \qquad  & \quad $f_1$ &  $f_2$ \quad \quad & \quad \quad $f_3$ & $f_4$\\ \hline
0 \qquad  & \quad $0$ &  $1$ \quad \quad  & \quad \quad $0$& $1$\\ \hline
1 \qquad  & \quad $0$ &  $1$ \quad \quad  & \quad \quad $1$& $0$\\ \hline
$U_f$\qquad  & \quad {\small $\pmatrix{1&0&0&0 \cr 0&1&0&0 \cr 0&0&1&0 \cr 0&0&0&1}$} &
{\small$\pmatrix{0&1&0&0 \cr 1&0&0&0 \cr 0&0&0&1 \cr 0&0&1&0}$}
\quad \quad  & \quad \quad
{\small $\pmatrix{1&0&0&0 \cr 0&1&0&0 \cr 0&0&0&1 \cr 0&0&1&0}$} &
{\small $\pmatrix{0&1&0&0 \cr 1&0&0&0 \cr 0&0&1&0 \cr 0&0&0&1}$}\\ \hline
Pulse \qquad  & \quad  $no~~ pulse$ &
$(\pi)^{\vert 00 \rangle \leftrightarrow \vert 01 \rangle}-(\pi)^{\vert 10 \rangle \leftrightarrow \vert 11 \rangle}$
\quad \quad  & \quad \quad
$(\pi)^{\vert 10 \rangle \leftrightarrow \vert 11 \rangle}$ & $(\pi)^{\vert 00 \rangle \leftrightarrow \vert 01 \rangle}$ \cr \hline
\end{tabular}
\end{center}
\vspace{1cm}
\begin{center}
 Table 2. Transition numbers of expression 10 and corresponding phases of the r.f. pulses for creating
$\vert 00 \rangle + \vert 11 \rangle$ EPR state.  The transition numbers are from Figure 1(b).\\
\begin{tabular}{|c|cc|cc||c|cc|cc|}
\hline
Expt.& $k$ & $l$ & $\phi_1$ & $\phi_2$ &Expt. & $k$ & $l$ & $\phi_1$ & $\phi_2$\\
No.& & &  &  &No. &  &  &  & \\
\hline
\hline
1 & 2 & 3 & x & -x & 5 & 4 & 1 & x & -x \\
2 & 2 & 3 & -x & x & 6 & 4 & 1 & -x & x \\
3 & 2 & 3 & y & y & 7 & 4 & 1 & y & y \\
4 & 2 & 3 & -y & -y & 8 & 4 & 1 & -y & -y \\
\hline
\end{tabular}
\end{center}
\vspace{1cm}
\begin{center}
 Table 3. The 16-step phase-cycle for preparing the GHZ state.  The pulse-sequence is given in expression \ref{ghzpulses}.\\
\end{center}
\begin{center}
\begin{tabular}{||c|rrr||c|rrr||c|rrr||c|rrr||}
\hline
& $\phi_1$ & $\phi_2$ & $\phi_3$ && $\phi_1$ & $\phi_2$ & $\phi_3$ &&
$\phi_1$ & $\phi_2$ & $\phi_3$ && $\phi_1$ & $\phi_2$ & $\phi_3$ \\
\hline
~1~ & y & y & y & ~5~ & y & x & -x &~ 9~ & x & y & -x &~ 13~ & x & x & -y \\
~2~ & y & -y & -y &~ 6~ & y & -x & x &~ 10~ & x & -y & x &~ 14~ & x & -x & y \\
~3~ & -y & y & -y &~ 7~ & -y & x & x &~ 11~ & -x & y & x &~ 15~ & -x & x & y \\
~4~ & -y & -y & y &~ 8~ & -y & -x & -x &~ 12~ & -x & -y & -x &~ 16~ & -x & -x & -y \\
\hline
\end{tabular}
\end{center}
\vspace{1cm}

\newpage
\begin{center}
{\bf FIGURE CAPTIONS}
\end{center}
Figure 1. (a) Eigenstates of a strongly coupled two-spin
system, (b) qubit labeling, (c) trisodiumcitrate yielding a strongly coupled two-spin
system, and (d) the equilibrium 500MHz 
$^1$H spectrum of (c). \\\\

Figure 2. (a) Equilibrium spectrum of trisodiumcitrate, and spectra corresponding to various pseudopure states
(b) $\vert 00 \rangle$, (c) $\vert 01 \rangle$, (d) $\vert 10 \rangle$ and (e) $\vert 11 \rangle$.  
 Numbers above the enrgy levels indicate populations and the binary numbers below the levels indicate the labels. 
Transitions are also indentified as 1, 2 , 3 and 4 in (a).
The pulse sequences applied to prepare each pseudopure state is explained in the text.   Transition selective
pulses used were of length 100 ms and the gradient pulse was of length 1 ms and strength 10 G/cm.  Each spectrum
was obtained by a non-selective high power pulse of duration 1 $\mu$s corresponding to a flip-angle 
of 10$^{\circ}$. \\\\

Figure 3. Implementation of Deutsch-Jozsa algorithm. (a) Quantum circuit of and (b) experimental scheme for the  
implementation of DJ algorithm.
(c) and (d) are the experimental spectra corresponding to the two constant functions of $U_1$ and $U_2$ respectively, and 
(e) and (f) corresponding to the two balanced functions of $U_3$ and $U_4$ respectively.
The expected spectrum for all the four functions are given on the right hand side.  \\\\

Figure 4. Creation and tomography of $(\vert 00 \rangle + \vert 11 \rangle)/\sqrt2$ EPR state.
Spectra in (a-e) are experimental and in (f-g) are simulated corresponding to various steps of creation
and tomography. (a,f) The equilibrium spectra, (b,g) spectra after creation EPR state
    (no single quantum signal observed); (c,h) the spectra of diagonal part measured by 
    the pulse sequence $(EPR)-G_z-10^{\circ}_x$; (d,i) spectra obtained by the sequence
    $(EPR)-(\pi/4)_x-10^\circ_x$, (e,j) spectra obtained by  $(EPR)-(\pi/4)_y-10^\circ_x$.
(k) The two dimensional spectrum to measure the off-diagonal elements.  This spectrum clearly shows the double quantum
    coherence present in the EPR state. The axial peaks at zero frequency originate from the 
 longitudinal relaxation of EPR state during $t_1$ period 
 which is detected due to the imperfection of the $90^o$ r.f. pulse following the $t_1$ period.
The theoretical (l) and, and experimental (m) density matrices of EPR state.
    Spectra in (d,e) are used for calculating the scaling between diagonal and off-diagonal measurements.
    While plotting, the spectra shown in (d,e,i,j) are scaled up by a factor of 4.  Pseudopure
    state and EPR state are created by using transition selective pulses of length 100 ms.  An eight step cycle (shown in
    Table 2) was employed to minimize the errors. \\\\

Figure 5. Pulse sequence for creation and tomography of EPR state.
Numbers inside the parenthesis indicate transition numbers as shown
in Figure 1(d).
${\sf U_{EPR}}$ is applied on transitions [(2),(3)] or
[(4),(1)] as described in the expression \ref{eprpulex} and
Table 2.
$G_1$, $G_2$ and $G_3$ are field gradient
pulses of different strengths along $\hat{z}$-direction. \\\\

Figure 6. (a) Organometallic compound (I) in which the two Phosphorus ($^{31}$P) nuclei constitute a two spin system. 
(b) energy level diagram of the two spin system and 
(c) the equilibrium phosphorus spectrum recorded at 202 MHz in a magnetic field of 11.7 Tesla. \\\\

Figure 7. Implementation of 24 one-to-one logic gates. Starting from equilibrium all the gates were implemented using 
sequences of transition selective pulses and non-selective pulses. The unitary transforms and
pulse sequences for implementation of these gates are given in reference \cite{mahesh01}. Gaussian shaped 
 pulses of 100ms duration were used as selective pulses. The r.f. power
has been calibrated for given angle of flip for the two inner versus the two outer transitions. A sine-bell shaped 
gradient was applied after implementation of each selective pulse, to kill any coherences created due to imperfection of pulses.  
The final populations were mapped by a small angle $(10^o)$ non-selective pulse. \\\\

Figure 8. Equilibrium proton spectrum of 3-bromo-1,2-dichlorobenzene oriented in 
ZLI-1132 at 500 MHz. The transitions are labeled from left to right.  The ninth
transition marked * did not show connectivity to other transitions (Figure 9), and belongs to the lone transition between 011 and 100 
(markedby dashed line in Figure 10). \\\\

Figure 9. Z-COSY spectrum of oriented 3-bromo-1,2-dichlorobenzene. The equilibrium spectrum and the cross-sections of the
Z-COSY spectrum are shown on the right-hand side. \\\\

Figure 10. Energy level diagram of oriented 3-bromo-1,2-dichlorobenzene constructed using the
Z-COSY spectrum shown in Figure 9. The transitions are labeled as in Figure 8. Only nine transitions are assigned.  The remaining
transitions are having very low intensity. \\\\

Figure 11. Preparation of pseudopure states on the three spin strongly coupled system of Figure 8.
Energy levels, transitions and representative deviation populations (numbers inside the circles) are shown on
the left hand side and the corresponding spectra are shown on the right hand side.
(a) Equilibrium deviation populations and the corresponding
spectrum, (b) deviation populations and spectrum obtained
after inverting the transition {\bf 6}, and (c) the deviation
populations and the spectrum obtained by subtracting (b) from
(a).  The deviation populations and the spectrum in (c)
correspond to the pair of pseudopure states:
$\vert 000 \rangle \langle 000 \vert -
\vert 001 \rangle \langle 001 \vert$. 
(d) Population distribution and spectrum corresponding to the state
$\vert {\sf GHZ} \rangle \langle {\sf GHZ} \vert- \vert 001 \rangle \langle 001 \vert$.
All spectra were recorded using a final small angle $(10^o)$
detection pulse to maintain linear response such that
the intensities are proportional to the population differences of the two involved levels only.\\\\

Figure 12. (a) Pulse sequence for creation and tomography of
$\vert {\sf GHZ} \rangle \langle {\sf GHZ} \vert
- \vert 001 \rangle \langle 001 \vert$ state.
Numbers inside the parenthesis indicate transition numbers as shown in Figure 8.
$G_1$ and $G_2$ are field gradient pulses of different strengths along $\hat{z}$-direction. (b) The two-dimensional spectrum obtained
by using the pulse sequence (a). Pure triple quantum coherence at $\omega_1+\omega_4+\omega_8$ (where, $\omega_k$ is the frequency
of the transition $k$) confirms the creation of GHZ state.\\\\

Figure 13. The result of DJ algorithm on 3-bromo-1,2-dichlorobenzene desolved in ZLI-1132 for various functions $f_1-f_8$. Only the
expansions of the transitions of the input qubits ($I_1$ and $I_2$) are shown. Expected patterns obtained by GAMMA simulation are also 
shown against each spectra. Transitions 3,4,9(*) and 6 are used as the four work qubit transitions. Interchange of labels 
011$\leftrightarrow$101 allows us to identify transitions 4 and 9 as transitions belonging to the 3rd qubit. This does not affect other 
operations and the four 3rd qubit transitions namely 3,4,9 and 6 remain unconnected. However transition 1 now belongs to 1st qubit 
(011$\leftrightarrow$111) along with 2 (010$\leftrightarrow$110) and transition 7 to 2nd qubit (001$\leftrightarrow$011) along with 8 
(000$\leftrightarrow$010). All the experiments were carried out on a Bruker DRX-500 spectrometer at 300K. The transition selective pulses 
were 1.5,7.4,20 and 1.5 ms long respectively for transitions 3,4,9 and 6. The pulse power was then adjusted to make the flip angle of each
pulse as $\pi$. A phase cycle of $(x,-x)$ was used to minimize the error of the $\pi$-pulses during computation. The extra peaks in
$f_2$ (shown by ``$\leftarrow$" marks) in the experimental spectra were originated due to undesired coherence transfer during computation.
All the experiments were done using 2048 $t_2$ and 128 $t_1$ data points. All plots are shown in magnitude mode. The resonance frequencies 
of various transitions (1,2,7 and 8) in $\omega_2$ domain are schematically identified in the bottom line of the Figure. The $\pi$ pulses
applied to various work qubit transitions are indicated for each f, with the transitions identified in $f_2$. The same order follows for 
other f's. For example, $f_2=(\pi,\pi,\pi,\pi)$ means $\pi$ pulses are applied to all the transitions of work qubit and $f_5=(\pi,0,\pi,0)$ 
means $\pi$ pulses are applied to transitions 3 and 9 and no pulses to transitions 4 and 6.\\\\

Figure 14. One dimensional 500 MHz proton spectrum of 2-chloroiodobenzene oriented in liquid crystal 
 ZLI-1132, at 300 K forming a 4-qubit system. 
 The transitions are labeled according to descending order of their intensities. \\\\

Figure 15. Z-COSY spectrum of the 4-spin strongly coupled system shown in Figure 13.
The spectrum consists of more than 2000 desired peaks. \\\\

Figure 16. (a) Labeling scheme for the states of the 4-qubit system of Figure 13.
(b) Equilibrium spectrum obtained using a small angle (10$^o$) pulse [the spectra of Fig. 13 was obtained using 
a 90$^o$ detection pulse].
(c) Spectrum corresponding to C$^3$-{\sf NOT} gate obtained by selective inversion of transition number 4 
($\vert 1110 \rangle \leftrightarrow \vert 1111 \rangle$). 
(d) Spectrum corresponding to the pair of pseudopure states $\vert 1111 \rangle \langle 1111 \vert
- \vert 1110 \rangle \langle 1110 \vert$ obtained by subtracting (c) from (b), named POPS(4).
(e) Spectrum corresponding to the pair of pseudopure states $\vert 1110 \rangle \langle 1110 \vert
- \vert 1010 \rangle \langle 1010 \vert$ obtained by inverting the transition 1 and subtracting the
obtained spectrum from the equilibrium spectrum (b), named POPS(1). (f) Spectrum corresponding the pair of
pseudopure states $\vert 1101 \rangle \langle 1101 \vert - \vert 1010 \rangle \langle 1010 \vert$
obtained by applying C$^2$-{\sf SWAP} gate on (e). The pulse sequence for C$^2$-{\sf SWAP} gate
is given in the expression \ref{ccswapex}. (g) Spectrum corresponding the pair of
pseudopure states  $\vert 1101 \rangle \langle 1101 \vert
- \vert 1010 \rangle \langle 1010 \vert$  obtained by inverting the transition 15
and subtracting the spectrum for the equilibrium spectrum (b), named POPS(15).
Spectra (f) and (g) match fairly well, indicating good implementation of the C$^2$-{\sf SWAP}.
All spectra were recorded using a final small angle $(10^o)$ detection pulse to maintain linear response such that
the intensities are proportional to the population differences of the two involved levels only.

\newpage
\begin{center}
\begin{figure}
\epsfig{file=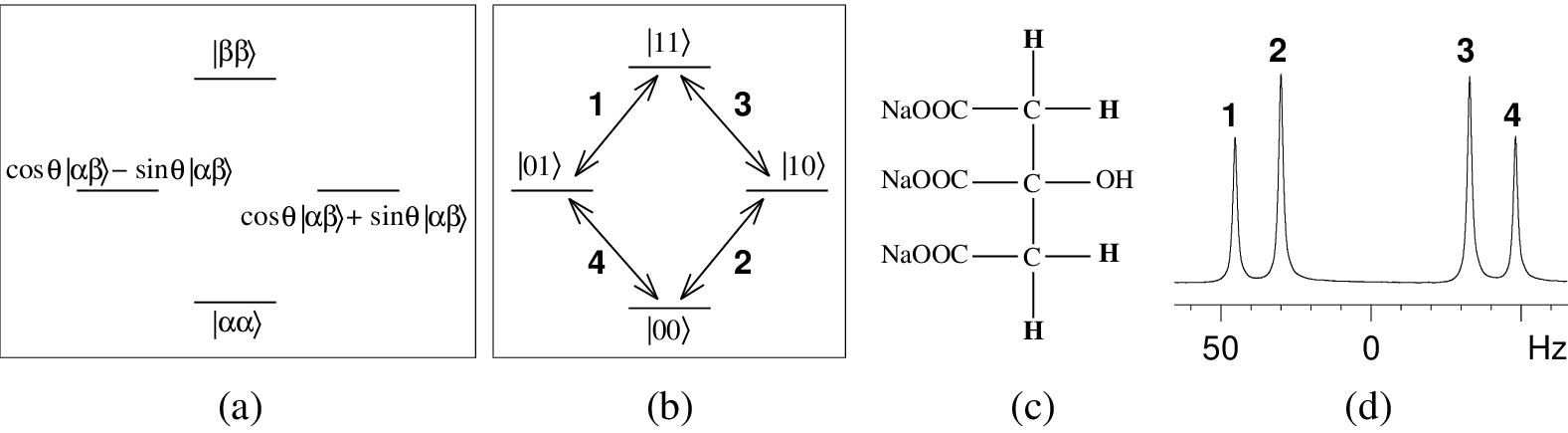,angle=-90,width=15cm}
\label{chp5enlevels}
\end{figure}
{\Large Figure 1}
\end{center}
\newpage
\vspace{2cm}
\begin{center}
\begin{figure}
\epsfig{file=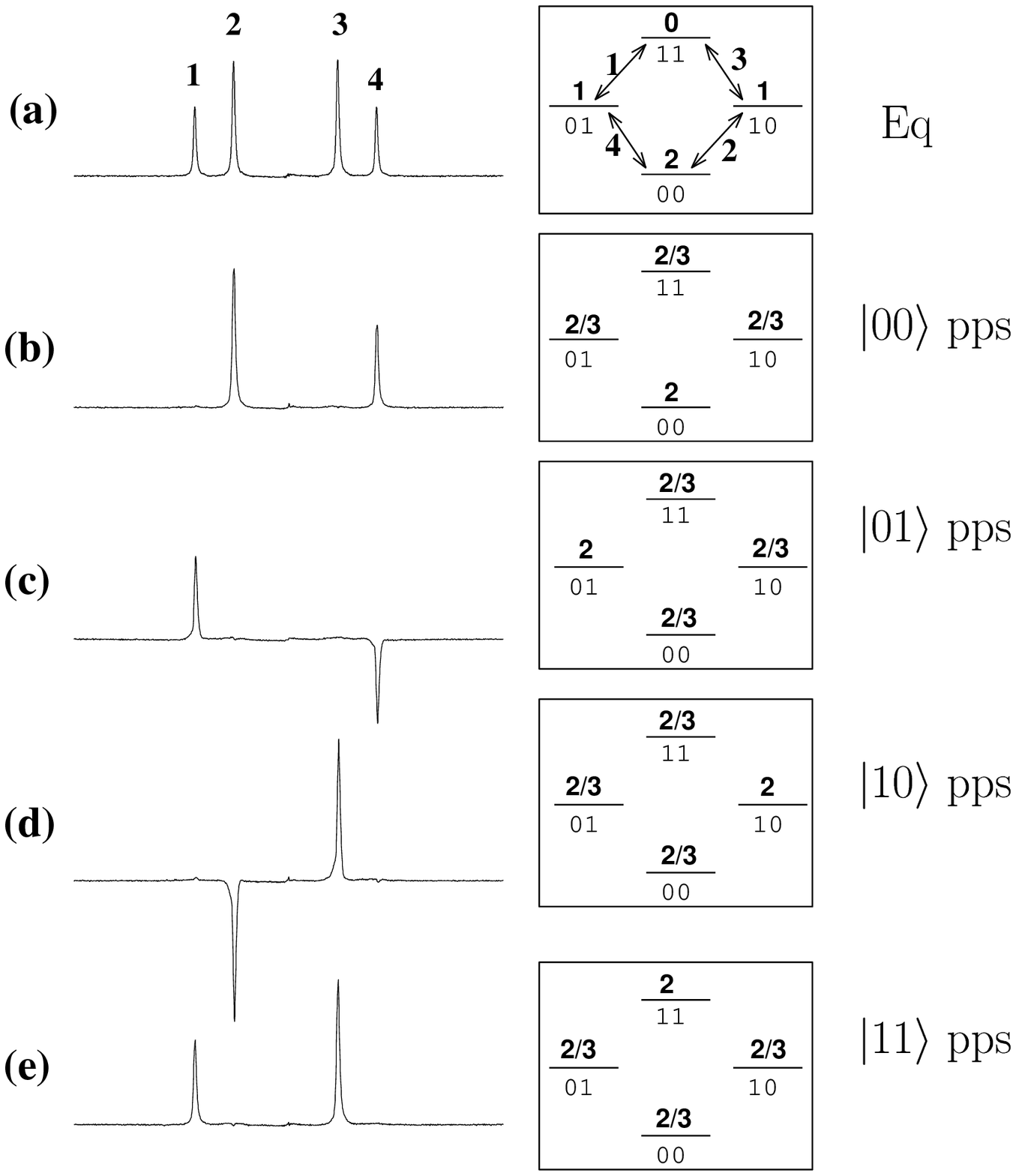,height=13cm}
\label{chp5pps}
\end{figure}
{\Large Figure 2}
\end{center}

\newpage
\begin{center}
\begin{figure}
\epsfig{file=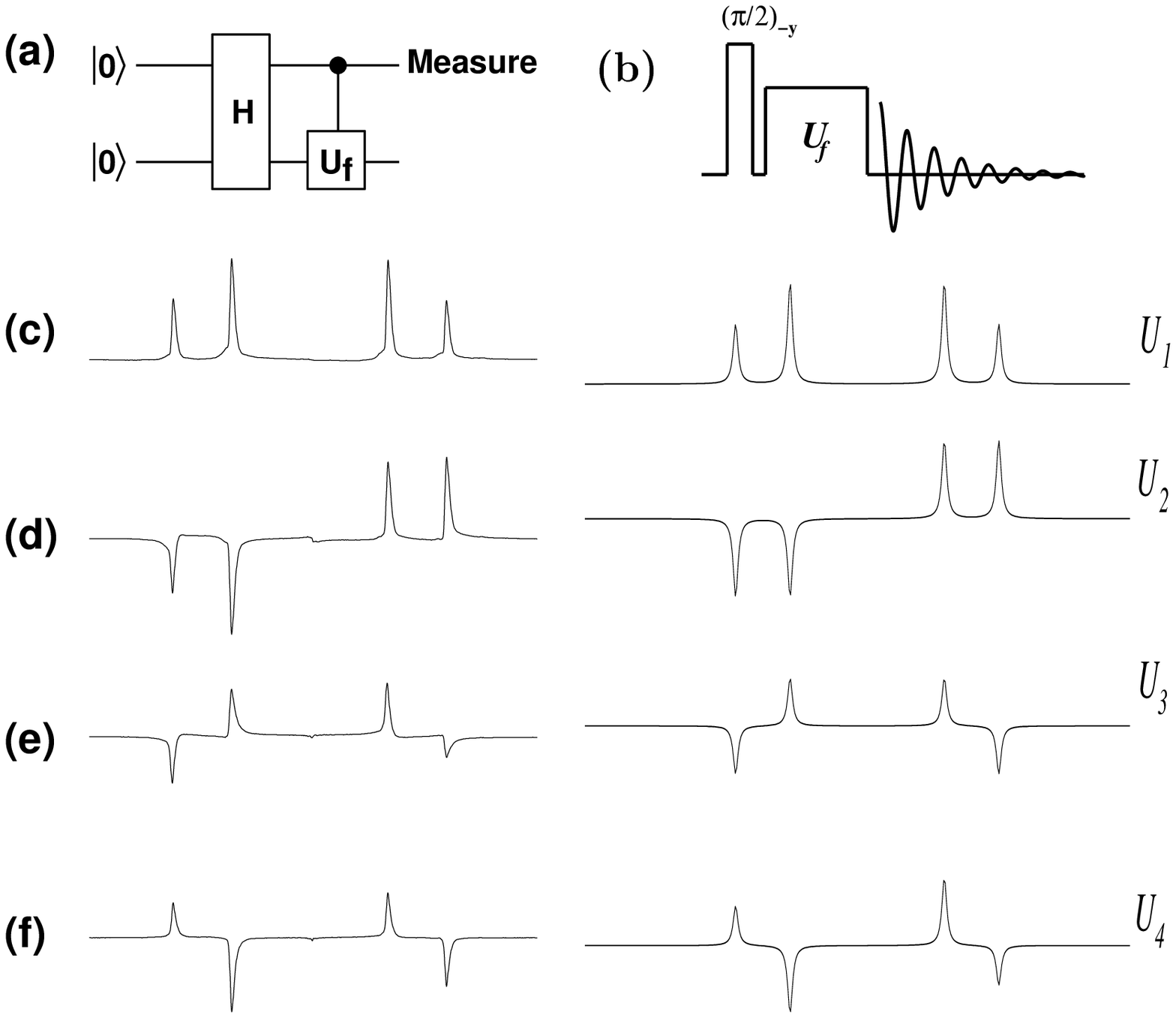}
\end{figure}
{\Large Figure 3}
\end{center}
\newpage
\begin{center}
\begin{figure}
\epsfig{file=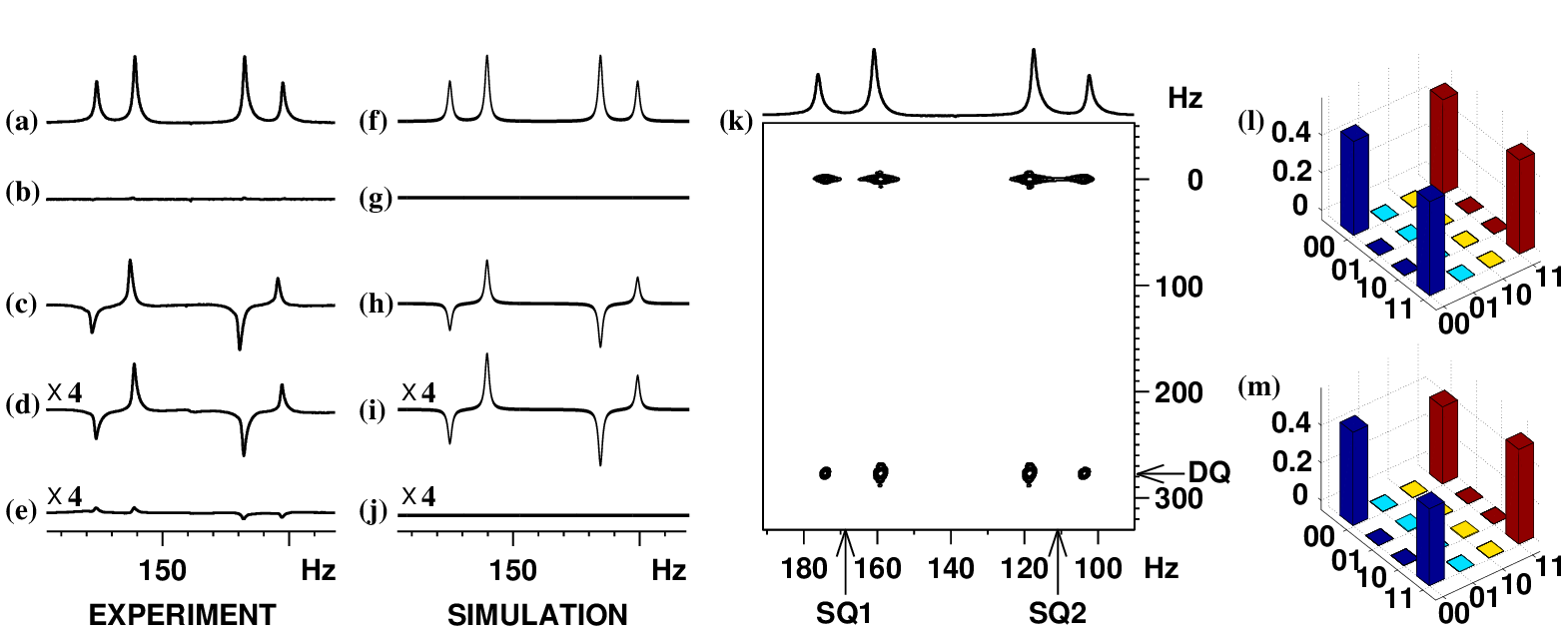,angle=-90,width=15cm}
\label{chp5epr}
\end{figure}
{\Large Figure 4}
\end{center}
\newpage
\vspace{3cm}
\begin{center}
\begin{figure}
\epsfig{file=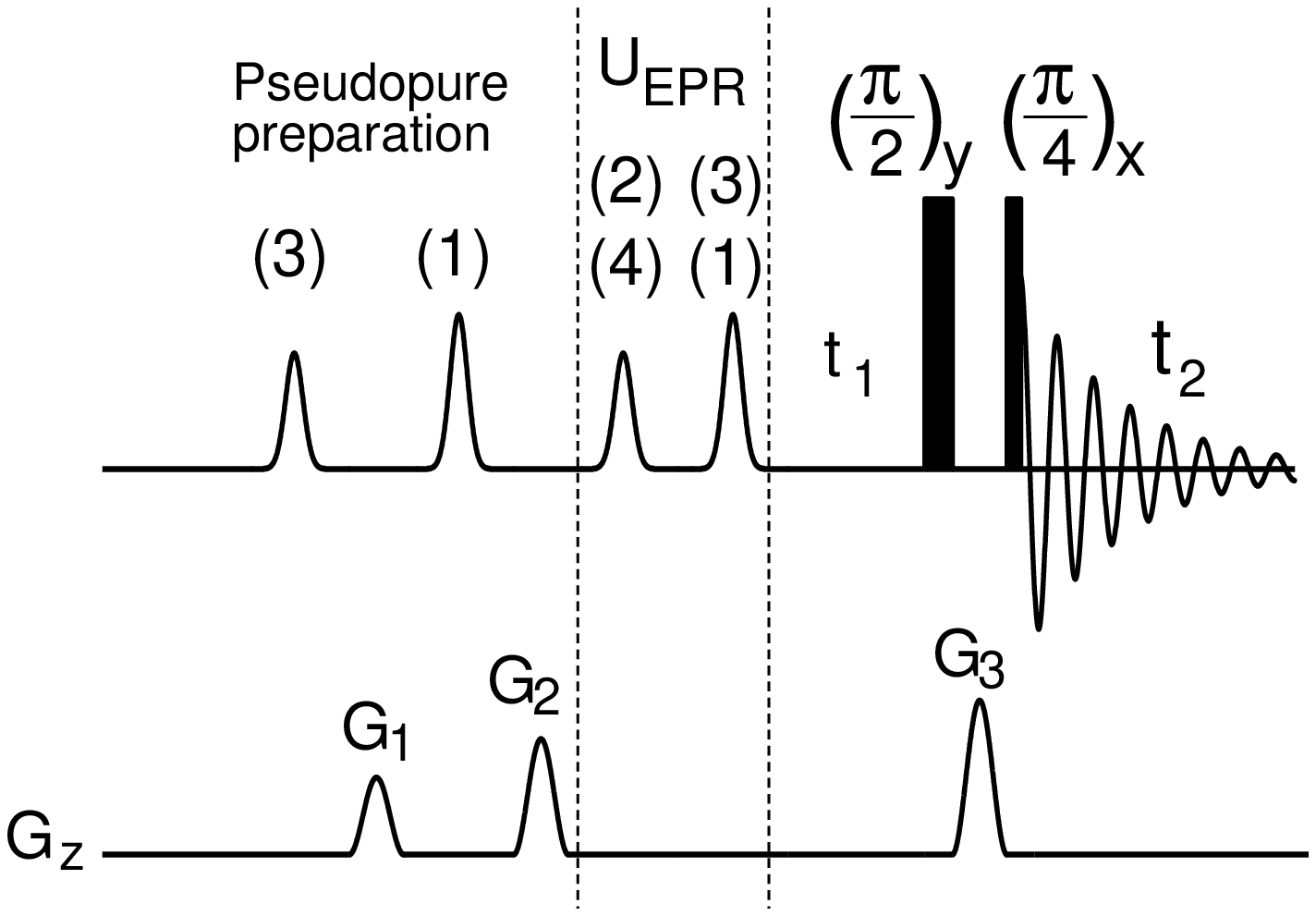,angle=-90,width=7cm}
\label{eprpulseq}
\end{figure}
{\Large Figure 5}
\end{center}
\newpage
\begin{center}
\begin{figure}
\epsfig{file=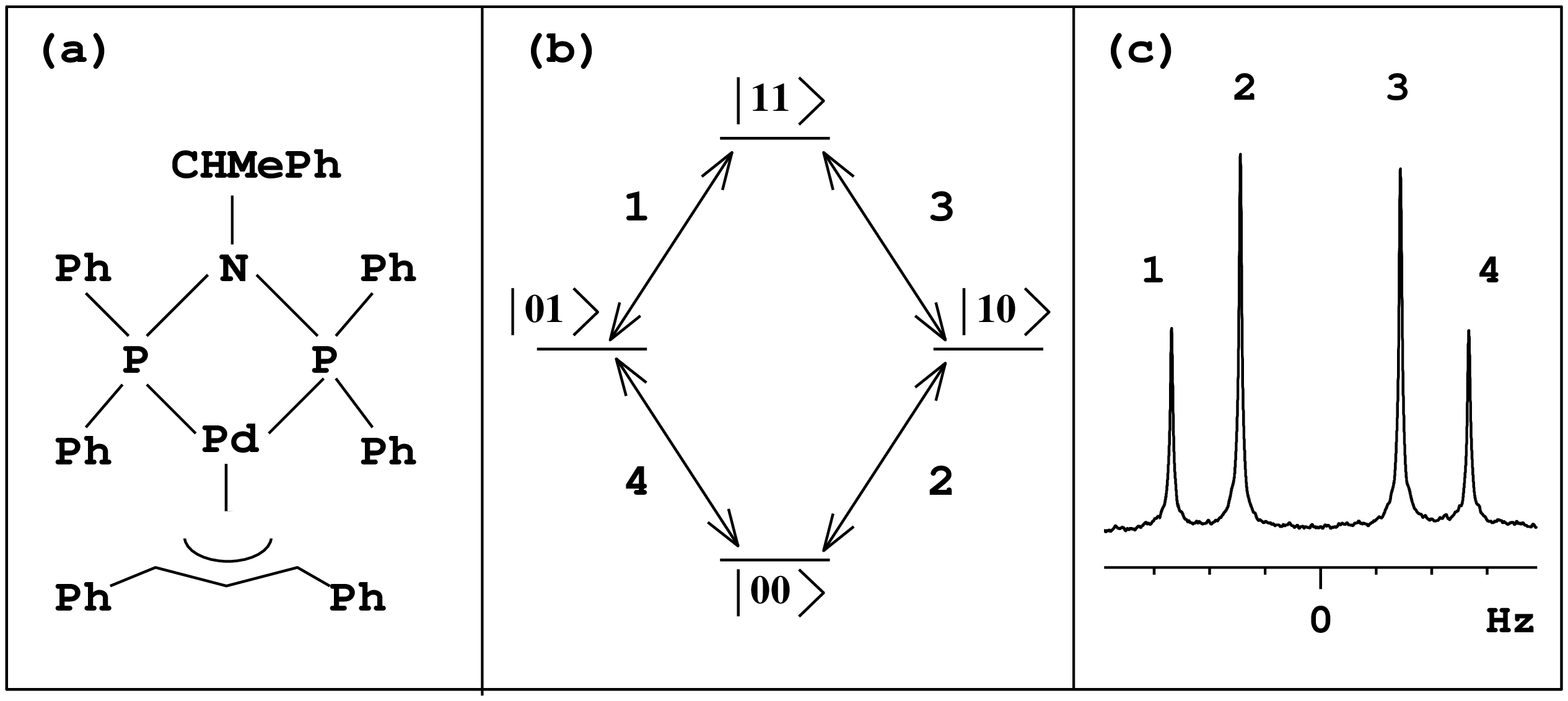,height=14cm,angle=270}
\end{figure}
{\Large Figure 6}
\end{center}
\newpage
\begin{center}
\begin{figure}
\epsfig{file=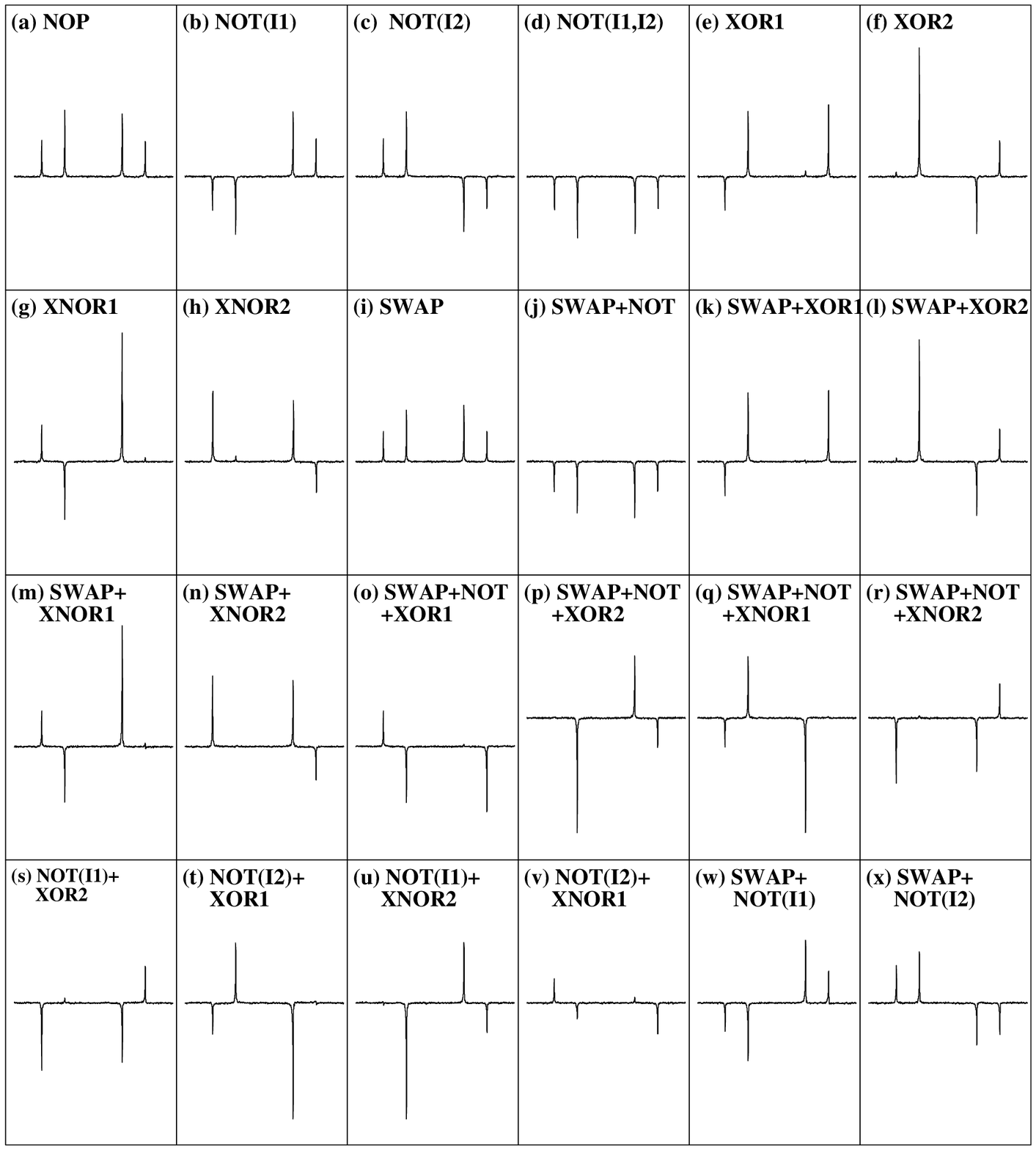,height=19cm}
\label{phos}
\end{figure}
{\Large Figure 7}
\end{center}
\newpage
\begin{center}
\begin{figure}
\epsfig{file=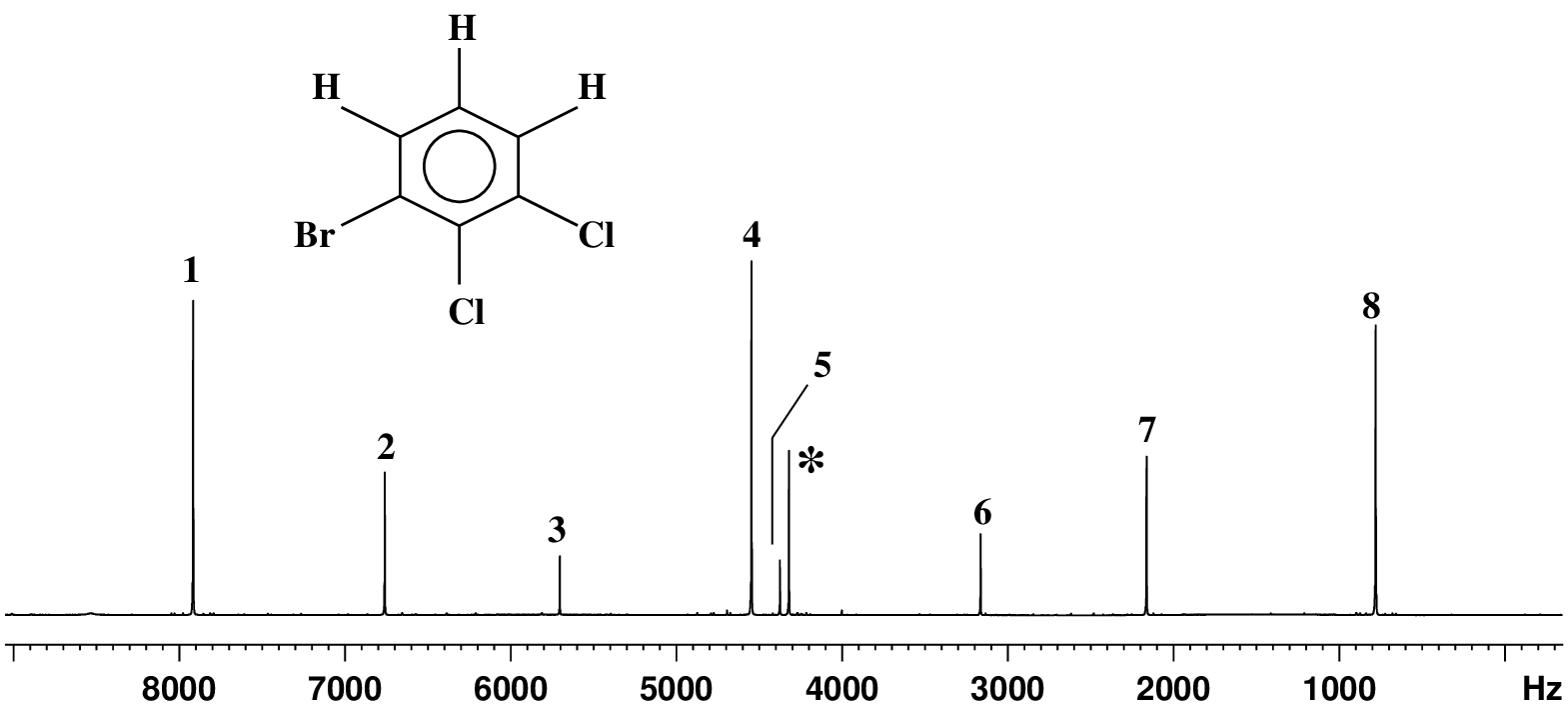,angle=-90,width=15cm}
\label{cl2br1d}
\end{figure}
{\Large Figure 8}
\end{center}
\newpage
\begin{center}
\begin{figure}
\epsfig{file=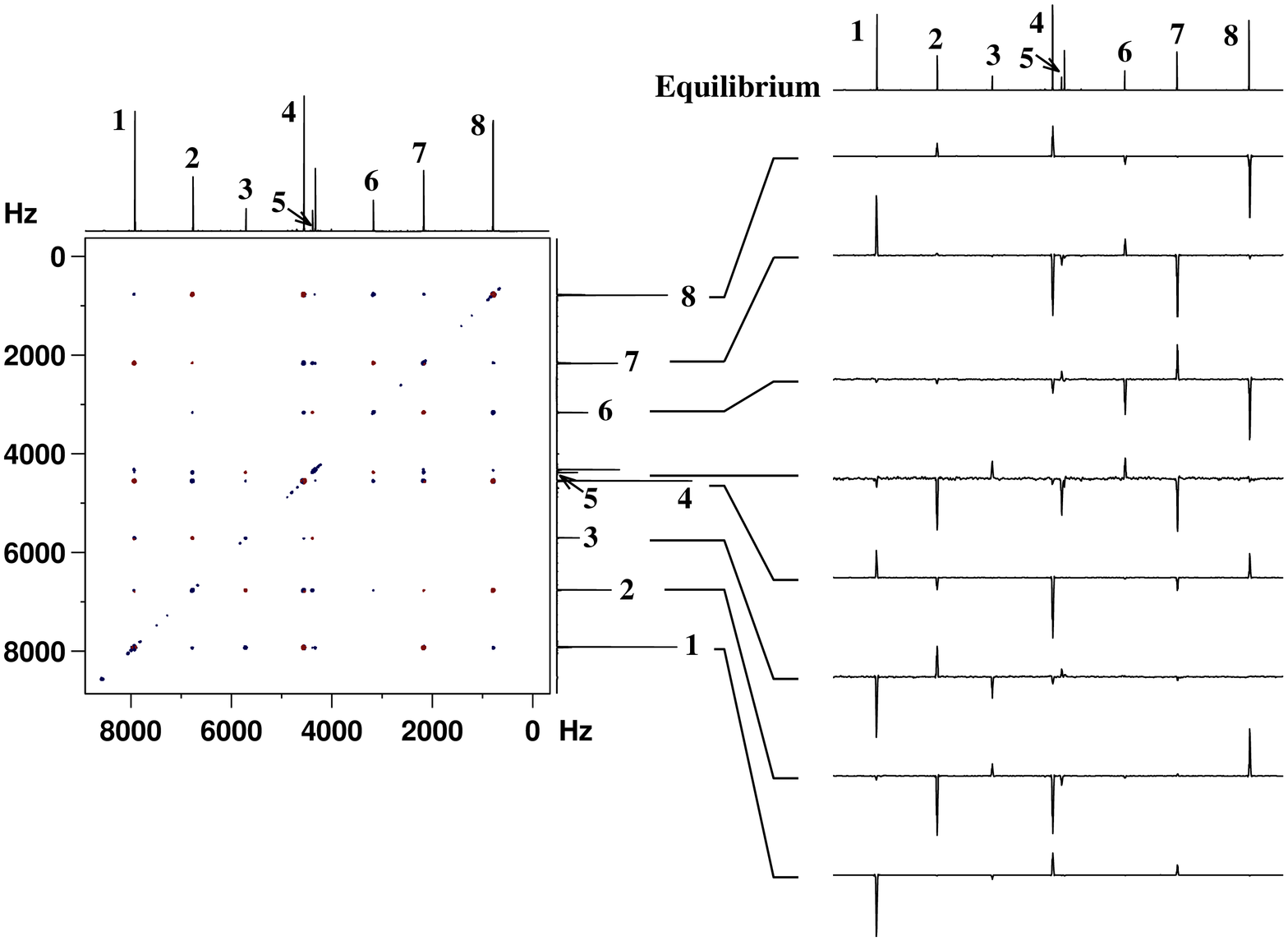,angle=-90,width=15cm}
\label{cl2brzcosy}
\end{figure}
{\Large Figure 9}
\end{center}
\newpage
\begin{center}
\begin{figure}
\epsfig{file=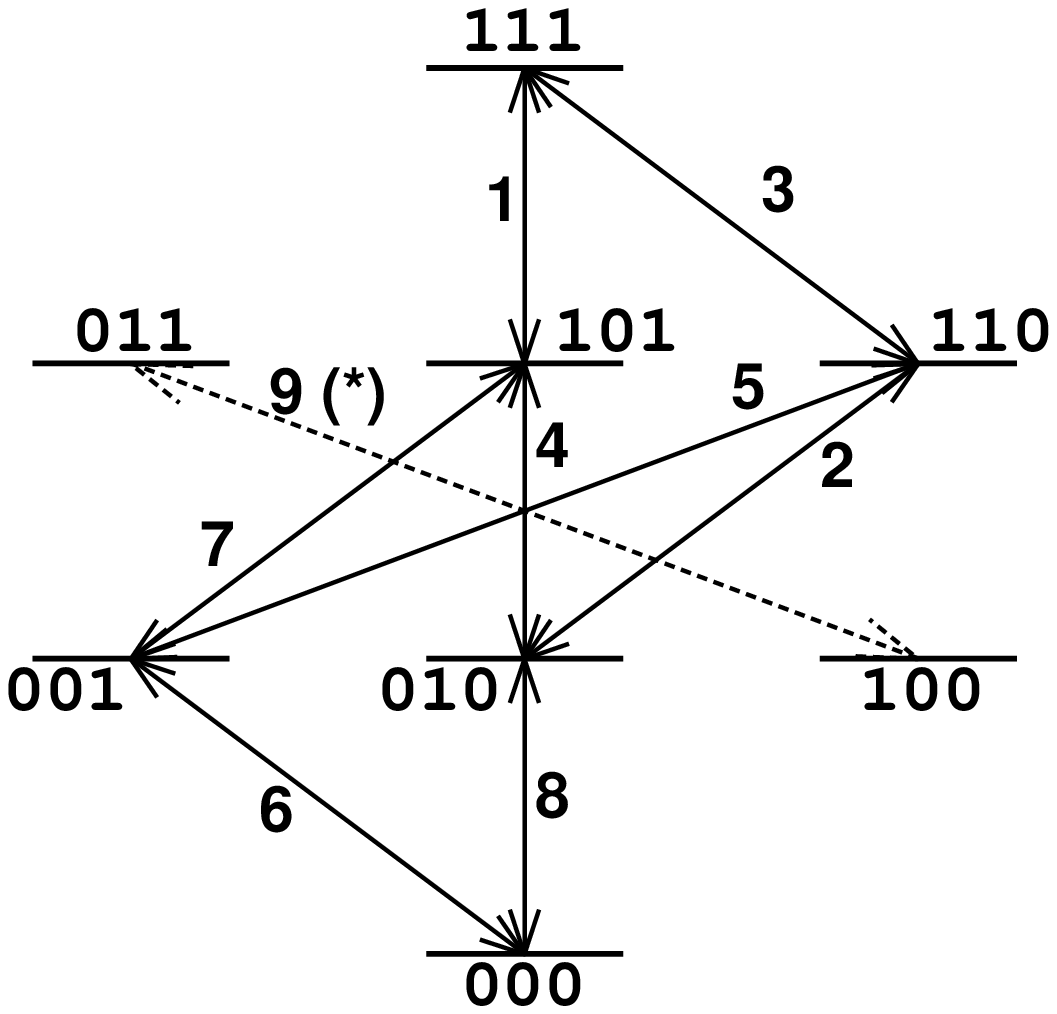,width=6cm,height=7cm}
\label{cl2brenlev}
\end{figure}
{\Large Figure 10}
\end{center}
\newpage
\vspace{1cm}
\begin{center}
\begin{figure}
\epsfig{file=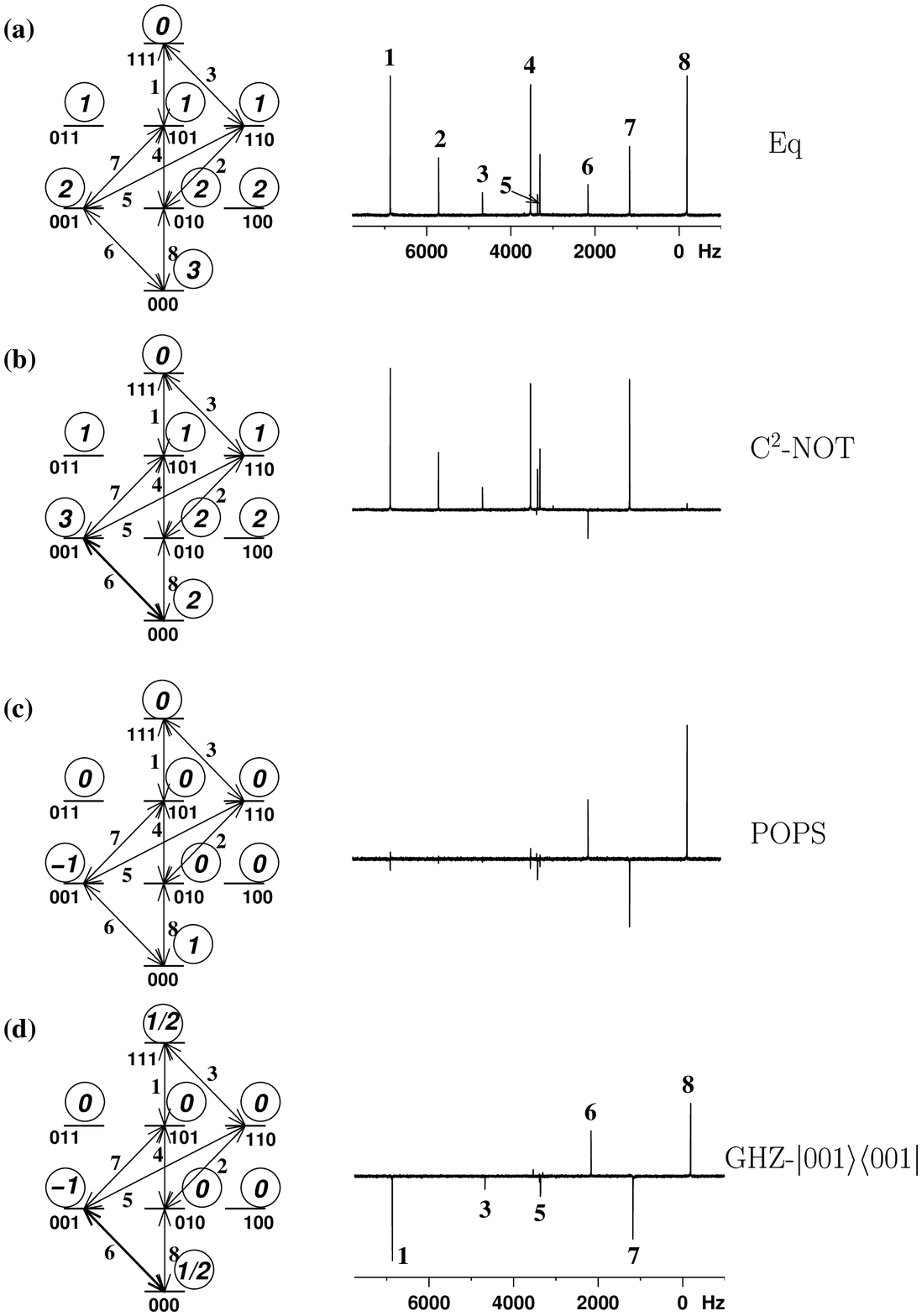,width=11cm}
\label{cl2brpps}
\end{figure}
{\Large Figure 11}
\end{center}
\newpage
\vspace*{-14cm}
\begin{center}
\begin{figure}
\epsfig{file=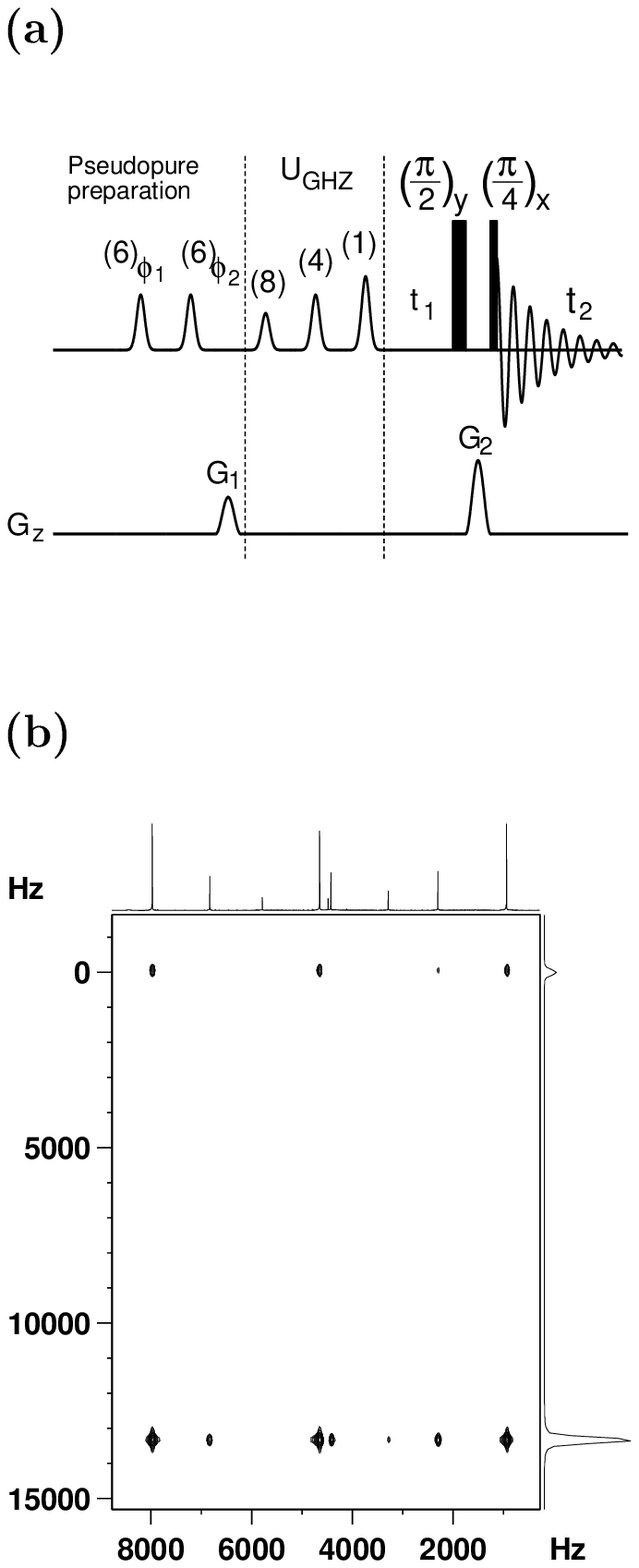}
\end{figure}
{\Large Figure 12}
\end{center}


\newpage
\begin{center}
\begin{figure}
\epsfig{file=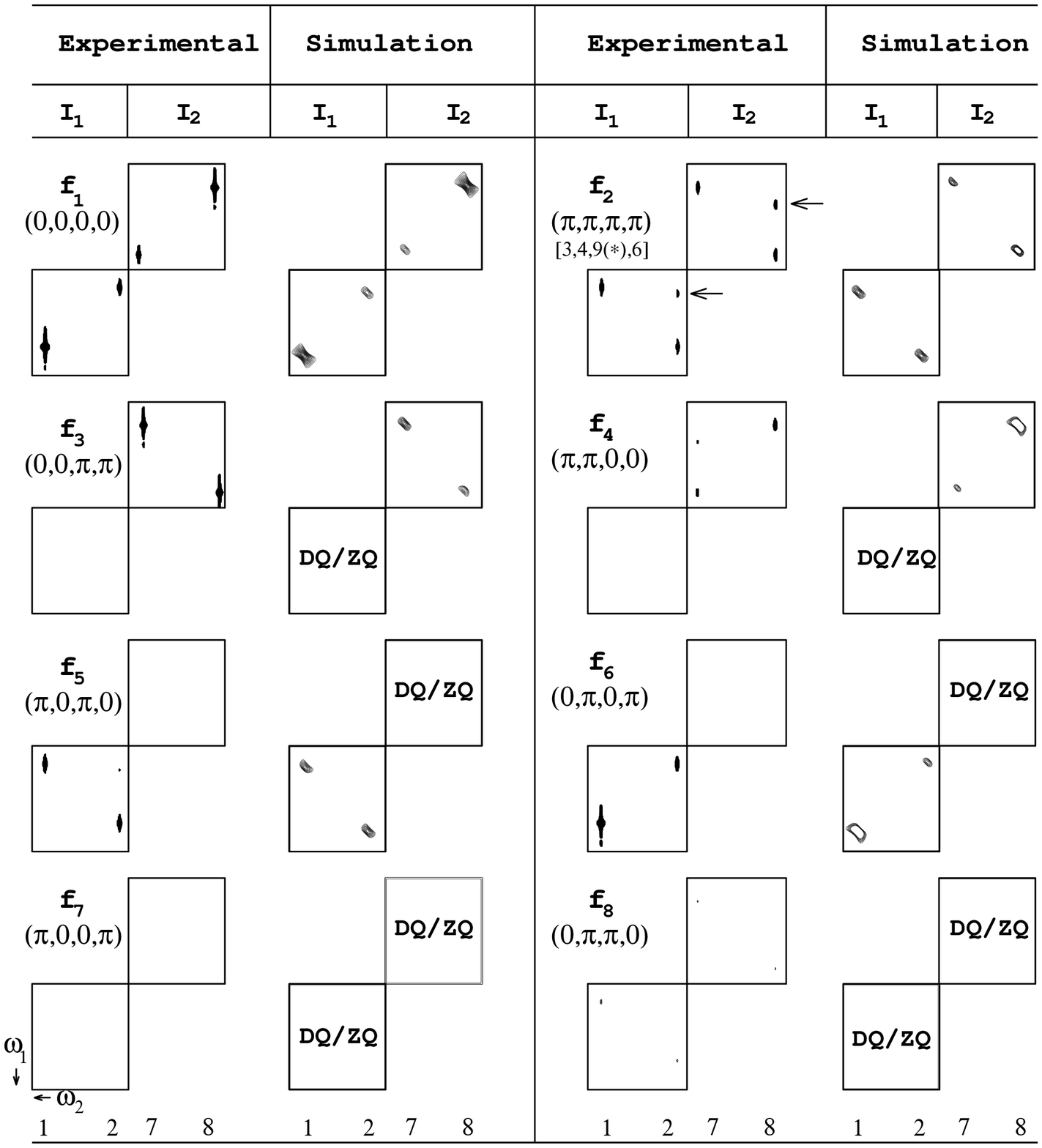,height=19cm}
\label{3qdj2d}
\end{figure}
{\Large Figure 13}
\end{center}
\newpage
\begin{center}
\begin{figure}
\epsfig{file=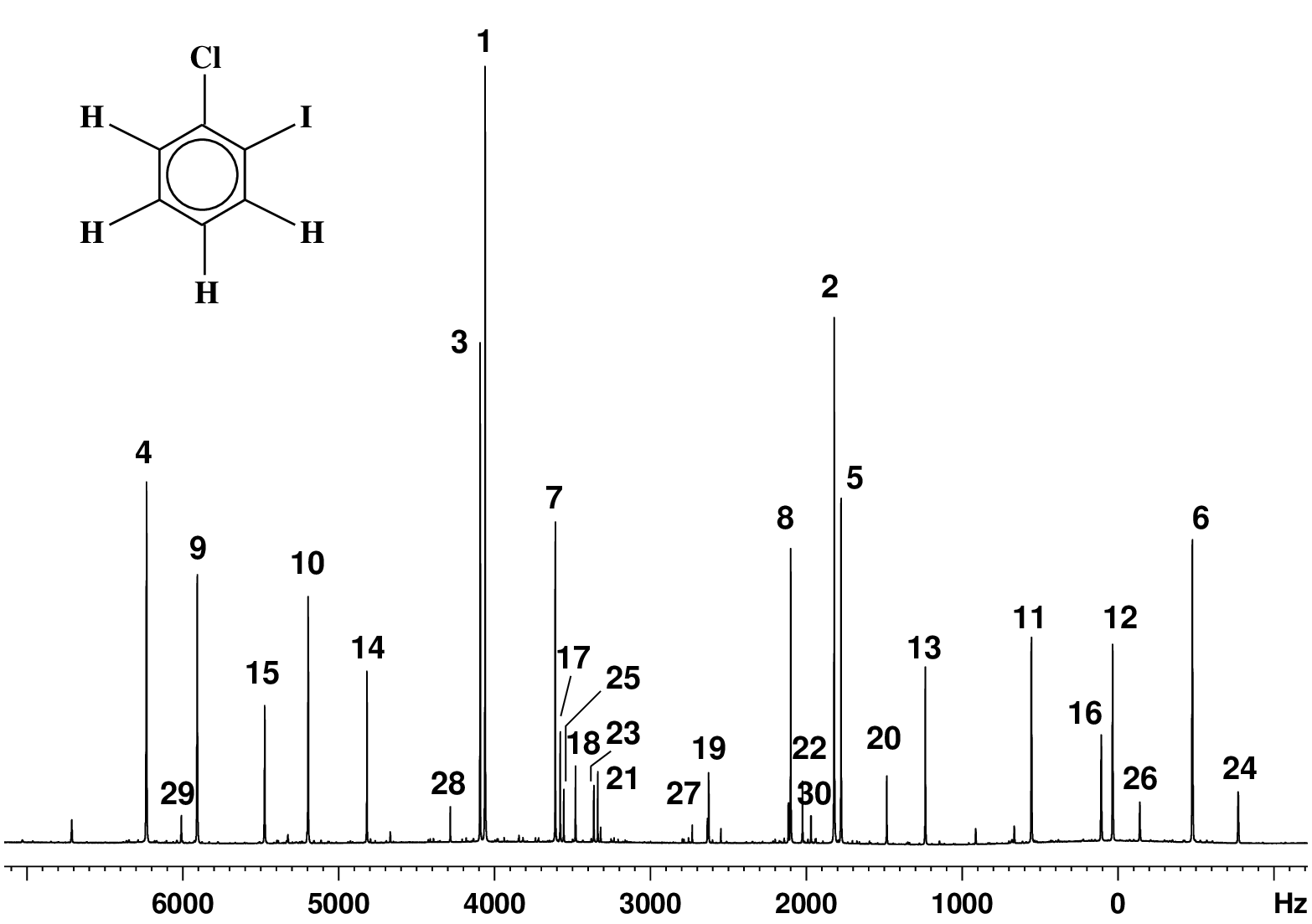,angle=-90,width=15cm}
\label{cli1dlabeling}
\end{figure}
{\Large Figure 14}
\end{center}
\newpage
\begin{center}
\begin{figure}
\epsfig{file=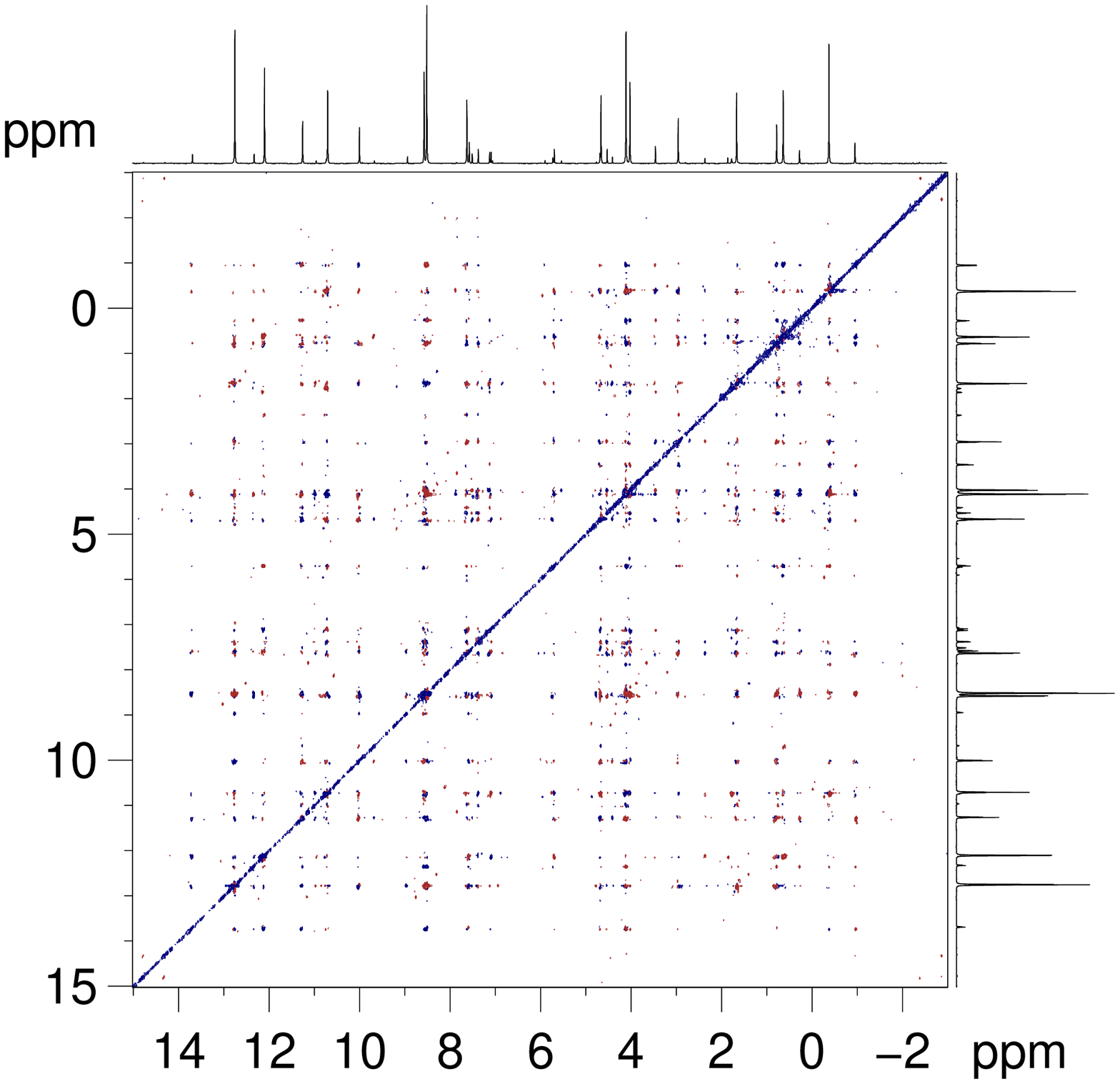,width=8cm,angle=-90}
\label{cli4qzcosy}
\end{figure}
{\Large Figure 15}
\end{center}
\newpage
\begin{figure}
\label{4qgatesstrong}
\hspace*{1.3cm}
\epsfig{file=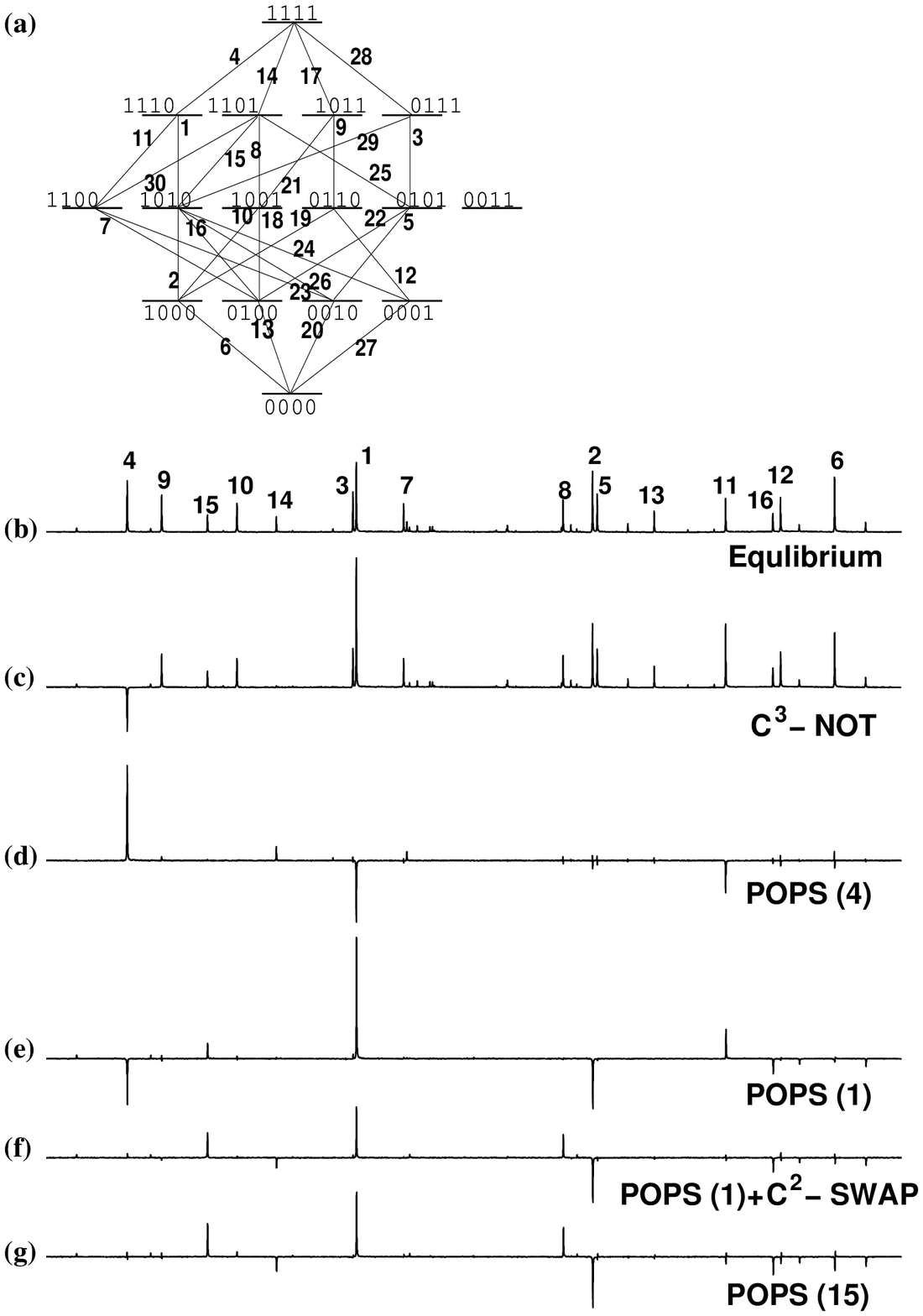,width=17cm}
\end{figure}
\begin{center}
{\Large Figure 16}
\end{center}


\end{document}